\documentclass[journal,comsoc]{IEEEtran}

\usepackage{afterpage}
\usepackage{amsmath}
\usepackage{amsthm}
\usepackage{balance}
\usepackage{booktabs}
\usepackage{breakcites}
\usepackage{caption}
\usepackage{cite}
\usepackage{fontawesome}
\usepackage[T1]{fontenc}
\usepackage{graphicx}
\usepackage[hidelinks]{hyperref}
\usepackage[utf8]{inputenc}
\usepackage{marvosym}
\usepackage{mathptmx}
\usepackage[cmintegrals]{newtxmath}
\usepackage{lscape} 
\usepackage{siunitx}
\usepackage{subcaption}
\usepackage{tcolorbox}
\usepackage{tikz}
\usepackage{wasysym}
\usepackage{xargs}

\newcounter{lessonctr}
\newcounter{openctr}

\newenvironment{lesson}{
    \refstepcounter{lessonctr}
    \begin{tcolorbox}[bottomrule=1mm, colback=white, colframe=black,arc=4mm,sharp corners=downhill]
        \textbf{Observation \thelessonctr:}
}{
    \end{tcolorbox}
}

\newenvironment{open}{
    \refstepcounter{openctr}
    \begin{tcolorbox}[bottomrule=1mm, colback=white, colframe=black,arc=4mm,sharp corners=downhill]
        \textbf{Open Question \theopenctr:}
}{
    \end{tcolorbox}
}

\usetikzlibrary{calc,shapes.geometric,arrows,positioning,arrows.meta}

\usepackage{xcolor}

\newcommand\secpar[1]{\noindent \textbf{#1:}}

\newcommand\email[1]{\href{mailto:#1}{#1}}

\newcommand\rot[0]{\rotatebox{90}}
\newcommand\yes[0]{\checked}
\newcommand\no[0]{--}
\newcommand\proposed[0]{\hexstar}

\newcommand\acoustic[0]{\faVolumeUp{}}
\newcommand\conducted[0]{\faBolt{}}
\newcommand\electromagnetic[0]{\faWifi{}}
\newcommand\optical[0]{{\small \Laserbeam{}}}

\usepackage{ifthen}

\newcommand{\pie}[1]{%
\begin{tikzpicture}
 \draw (0,0) circle (1ex);\ifthenelse{\equal{#1}{0}}
 {}
 {\fill (1ex,0) arc (0:#1:1ex) -- (0,0) -- cycle;}
\end{tikzpicture}%
}

\newcommand\theoretical[0]{\pie{0}}
\newcommand\bias[0]{\rotatebox{90}{\pie{180}}}
\newcommand\disruption[0]{\rotatebox{90}{\pie{90}}}
\newcommand\control[0]{\pie{360}}

\newcommandx{\gpath}[4][1=0, 4=thick]{\path(#2) edge [-{latex[scale=2.5]}, #4, bend left=#1] (#3);}

\graphicspath{{./figures/}}

\newtheorem{definition}{Definition}

\begin{document}

\pagenumbering{gobble}
\onecolumn
\begin{center}
{\Large Taxonomy and Challenges of Out-of-Band Signal Injection Attacks and Defenses}\\
~\\
{\large Ilias Giechaskiel, Kasper B. Rasmussen}\\
~\\
{in {\em IEEE Communications Surveys \& Tutorials}, vol. 22, no. 1, pp. 645--670, Firstquarter 2020.}
\end{center}

\vspace*{\fill}
\noindent
\copyright 2020 IEEE. Personal use of this material is permitted. Permission from IEEE must be obtained for all other uses, in any current or future media, including reprinting/republishing this material for advertising or promotional purposes, creating new collective works, for resale or redistribution to servers or lists, or reuse of any copyrighted component of this work in other
works. DOI:\href{http://doi.org/10.1109/COMST.2019.2952858}{10.1109/COMST.2019.2952858}
\vspace*{\fill}
\clearpage
\pagenumbering{arabic}
\twocolumn

\title{Taxonomy and Challenges of Out-of-Band Signal Injection Attacks and Defenses}

\author{Ilias Giechaskiel and Kasper B. Rasmussen\vspace*{-2em}
\thanks{Manuscript received April 29, 2019; revised August 30, 2019; accepted November 03, 2019.
        Date of publication November 12, 2019; date of current version November 12, 2019.
        {Corresponding author: Ilias Giechaskiel.}}
\thanks{The  authors  are  with  the  Department  of  Computer Science, University of
Oxford, Oxford, UK  (e-mail: \email{ilias.giechaskiel@cs.ox.ac.uk};
\email{kasper.rasmussen@cs.ox.ac.uk}).}
\thanks{Digital Object Identifier 10.1109/COMST.2019.2952858}
}

\maketitle

\begin{abstract}
Recent research has shown that the integrity of sensor measurements can be violated
through out-of-band signal injection attacks. These attacks target the conversion
process from a physical quantity to an analog property---a process that fundamentally cannot
be authenticated. Out-of-band signal injection attacks thus pose previously-unexplored
security risks by exploiting hardware imperfections in the sensors themselves,
or in their interfaces to microcontrollers. In response to the growing-yet-disjointed
literature in the subject, this article presents the first survey of out-of-band
signal injection attacks. It focuses on unifying their terminology and identifying
commonalities in their causes and effects through a chronological, evolutionary,
and thematic taxonomy of attacks. By highlighting cross-influences between
different types of out-of-band signal injections, this paper underscores
the need for a common language irrespective of the attack method. By placing
attack and defense mechanisms in the wider context of their dual counterparts
of side-channel leakage and electromagnetic interference,
this study identifies common threads and gaps that can help guide and
inform future research. Overall, the ever-increasing reliance
on sensors embedded in everyday commodity devices necessitates that a
stronger focus be placed on improving the security of such systems against
out-of-band signal injection attacks.
\end{abstract}


\section{Introduction}
\label{sec:intro}
\IEEEPARstart{M}{athematically} secure algorithms can be broken in
practice due to a mismatch between the high-level system model
used for analysis and the real-world environment on which
code runs. For example, data-dependent electromagnetic, optical, and acoustic
emanations, as well as variations in power consumption can reveal the
information processed by a device~\cite{spreitzer_systematic_2018},
with or without the help of intentional faults~\cite{barenghi_fault_2012,
karaklajic_hardware_2013, yuce_fault_2018}. However, attacks exploiting hardware
imperfections are not limited to side-channel leakage of confidential data:
recent research has shown that it is possible to target the integrity of
sensor measurements in a similar {\em out-of-band} fashion.

These out-of-band signal injection attacks can be performed using
electromagnetic radiation exploiting
circuits unintentionally acting as receiver antennas~\cite{kune_ghost_2013, shoukry_non_invasive_2013,
selvaraj_induction_embedded_2018}, as well as optical~\cite{park_pump_2016,
yan_autonomous_vehicles_2016} and acoustic~\cite{son_rocking_2015, zhang_dolphin_2017,
bolton_blue_2018} emissions targeting flaws in the conversion process from
physical properties into electrical ones.
The systems attacked have been equally diverse, and include medical
devices~\cite{kune_ghost_2013}, drones~\cite{son_rocking_2015, trippel_walnut_2017},
hard drives~\cite{bolton_blue_2018} and cameras~\cite{petit_remote_2015}, among others.
However, despite the wide range of attack methods and devices targeted, research in the
field thus far has been disjointed.

The ad-hoc nature of this type of research might stem, in part, from the fact
that out-of-band signal injection attacks have so far only been (openly)
conducted in a lab environment. However, out-of-band attacks have still
garnered the interest of technological and mainstream
news publications outside of the academic community~\cite{brewster_want_2017,
chirgwin_boffins_2017, gallagher_sounds_2017, markoff_possible_2017,
wagenseil_sonic_2017, moore_sonic_2018}. They have even prompted
national agencies to issue Computer Emergency Readiness Team (CERT)
advisories~\cite{cisa_ics_2017}. In some cases, the effects of out-of-band
attacks can be fatal: for instance, Foo Kune et al.\ have demonstrated that
low-power attacker signals can trick cardiac implantable electrical devices into causing
pacing inhibition and defibrillation shocks~\cite{kune_ghost_2013}.
Although the techniques used are not identical to those of mass-produced sonic
repellents~\cite{winberg_can_2019} or commercial~\cite{bbc_dewsbury_2019}
and military~\cite{youssef_us_2019} jammers, out-of-band signal injection attacks
share the same potential for weaponization. As a result,
to bring attention to these potentially severe
issues, and to help designers better protect future hardware devices, this study
conducts the first comprehensive survey of out-of-band signal injection attacks.

\subsection{Survey Scope}
\label{sec:scope}

This article focuses on out-of-band signal injection attacks, which
target the connections between sensors, actuators, and microcontrollers, or exploit
imperfections in the hardware itself. Although the term is defined precisely in
Section~\ref{sec:terminology}, we note here a few key features of
the attacks investigated in this survey. The first property of out-of-band signal
injection attacks is that they aim to {\em change} values processed by a system, rather than
infer them. This fact distinguishes them from the
side-channel~\cite{spreitzer_systematic_2018} and fault-injection
attacks~\cite{barenghi_fault_2012, karaklajic_hardware_2013,
yuce_fault_2018} mentioned in the introduction.

The second feature is that out-of-band attacks do not change the measured
quantity itself. For instance, using electromagnetic signals
to change the audio recorded by a microphone~\cite{kune_ghost_2013} is an
example of an out-of-band signal injection attack, but heating a
temperature sensor with an open flame is not.

The final characteristic highlights the physical aspect of out-of-band signal
injection attacks. In other words, the attacks studied in this paper alter sensor measurements
or actuator inputs at the hardware layer instead of the protocol layer.
As a result, spoofing attacks of unauthenticated, digital communication interfaces
are out-of-scope. For example, wireless transmissions can be used to spoof the
pressure of car tires and trigger warning lights~\cite{rouf_security_2010},
alter the flow of insulin injections~\cite{li_hijacking_2011}, or change
pacemaker settings to deliver shock commands with implantable
cardiac defibrillators~\cite{halperin_pacemakers_2008}. However, because these
interfaces can easily be protected with cryptography, they are not
considered in this survey. For similar reasons, jamming, e.g., train signal
controls~\cite{lakshminarayana_signal_2018}, or over-powering legitimate
GPS signals~\cite{zeng_gps_2018} are out-of-scope because they rely
on intentional communication interfaces. As a result, they do not exploit
hardware imperfections, instead relying on high-power, in-band electromagnetic
transmissions. Finally, relay attacks on LiDARs, radars, and
sonars~\cite{chauhan_attack_2014, petit_remote_2015, yan_autonomous_vehicles_2016,
shin_illusion_2017, xu_analyzing_2018} are also not out-of-band signal injection attacks. This is
because, in order to spoof the distance between the attacker and the victim, they
depend on winning a race between the adversarial signal and the true reflected pulses.

That said, these related research areas can offer invaluable insight into
novel injection techniques and possible countermeasures. As a result, cross-disciplinary
connections are made throughout this work, but with a clear focus on how
they impact out-of-band signal injection attacks and defenses.

\subsection{Contributions \& Organization}
\label{sec:contributions}

Despite the growing literature in the area and extensive parallels between
prior work in hardware security research and out-of-band signal injection
attacks and defenses, no other article has made these
connections explicit, or traced their evolution through time, theme, or approach.
Our survey fills these gaps through the following contributions:

\begin{enumerate}
    \item   It unifies the diverse terminology used by different
            works (Section~\ref{sec:terminology}) and summarizes the
            threat model (Section~\ref{sec:model}) to create a
            common language through which to discuss attack and defense mechanisms.
    \item   It proposes the first chronological and thematic evolution of
            out-of-band signal injection attacks (Section~\ref{sec:taxonomy}), which
            highlights cross-influences between electromagnetic (Section~\ref{sec:em}),
            conducted (Section~\ref{sec:conducted}), acoustic (Section~\ref{sec:acoustic}),
            and other (Section~\ref{sec:others}) attacks.
    \item   It creates a taxonomy of countermeasures introduced to prevent and
            detect out-of-band attacks (Section~\ref{sec:defenses}).
    \item   It places attacks and defenses in the wider context of side-channel leakage and
            electromagnetic interference
            attacks (Section~\ref{sec:related}). Using these insights,
            this study identifies gaps in the experimental approach
            of published research, and proposes concrete steps to overcome these
            challenges in the future (Section~\ref{sec:future}).
\end{enumerate}

\section{Choice of Terminology}
\label{sec:terminology}

The terms used to describe the numerous acoustic, electromagnetic,
and optical attacks on sensor-to-microcontroller and microcontroller-to-actuator interfaces
have so far been inconsistent, with some works not even naming the attacks at
all~\cite{rasmussen_proximity_based_2009, wang_sonic_2017}.
This section sets out to identify and unify the nomenclature used as a first step towards
providing a common language through which to compare the various works. As
the threat model (Section~\ref{sec:model}) and the causes of vulnerability
(Section~\ref{sec:taxonomy}) will reveal, the commonalities in the attack techniques
highlight a need for an all-encompassing term irrespective of the
method of injection. In other words, although some of the more restrictive terms
are appropriate for describing specific attacks, we find that doing so can
can hide potential insights that arise from considering different types of attacks jointly.
The term we have chosen for this unification is {\em out-of-band signal injection attacks}:

\begin{definition}[Out-of-Band Signal Injection Attacks]
Out-of-band signal injection attacks are adversarial manipulations
of interfaces not intended for communication involving sensors/actuators
that cause a mismatch between the true physical property being
measured/acted upon and its digitized version.
\end{definition}

To motivate our definition, the term {\em injection} was chosen because it captures
the fact that values reported by a system are altered; it is not channel-specific;
and it has already been adopted by different works~\cite{markettos_trng_injection_2009,
bayon_trng_em_2012, buchovecka_frequency_2013, kune_ghost_2013, kasmi_iemi_2015,
martin_fault_2015, trippel_walnut_2017, zhang_dolphin_2017, esteves_remote_2018,
nashimoto_sensor_2018, osuka_em_2019, tu_injected_actuation_2018, giechaskiel_framework_2019}.
The {\em out-of-band} qualifier is necessary to distinguish the attacks studied in this survey
from signal injection attacks on sensors using pulse reflections
such as LiDARs~\cite{petit_remote_2015, yan_autonomous_vehicles_2016,
shin_illusion_2017}, signal injection attacks on the physical layer
of communication protocols~\cite{jin_physical_2015}, and false data injection
attacks~\cite{liu_grids_2009, lakshminarayana_signal_2018}. As explained in
Section~\ref{sec:scope}, these attacks are out-of-scope, as they do not depend
on hardware vulnerabilities, but instead use external communication interfaces.

By contrast, our term captures attacks which target interfaces using signals outside
of their intended frequency of operation. It includes ultraviolet or
infrared light against cameras which should only be recording the visible part
of the spectrum, and ultrasonic injections against microphones meant to be recording
only audible sounds: these attacks transmit signals that are literally outside the
operational band. It also includes electromagnetic signals
against systems without any (intentional) antennas, and acoustic attacks against gyroscopes and accelerometers: these are also out-of-band, since they inject signals through
channels other than the ones used by the sensor to measure the physical property.
Our use of the out-of-band modifier is therefore consistent with the definition
for out-of-band covert communication~\cite{carrara_covert_2016}. It has also
recently been used by Tu et al.~\cite{tu_injected_actuation_2018} to describe
acoustic attacks on inertial sensors, further motivating its choice in this survey.

It should be noted that earlier work~\cite{shin_sampling_2016, shin_illusion_2017}
has proposed a subdivision of signal injections attacks into
{\em regular-channel attacks}, which target the sensor structure itself
by ``using the same type of physical quantity sensed'', {\em transmission-channel
attacks}, which target the connection between the sensor output and the measurement
setup, and {\em side-channel attacks}, where the sensors themselves are
targeted, but ``by physical stimuli other than those they are supposed to
sense''. We do not adopt this categorization, as it generally follows the medium of injection
(optical, electromagnetic, and acoustic respectively). Moreover, regular-channel attacks
are usually in-band, while side-channel attacks have an overloaded meaning.

We similarly find {\em (intentional) interference}~\cite{bayon_trng_em_2012,
kune_ghost_2013, kasmi_iemi_2015, bayon_fault_2016,
esteves_remote_2018, osuka_em_2019, son_rocking_2015, bolton_blue_2018}
to be unsuitable as a term because: (a) it does not make it clear that the attackers
can in some cases inject waveforms of their choosing; and (b)
Intentional Electromagnetic Interference
(IEMI) has an established meaning in Electromagnetic Compatibility (EMC)
literature~\cite{radasky_introduction_2004, savage_overview_2012}. As
Section~\ref{sec:related} indicates, IEMI attacks often use high-power,
destructive transmissions, and therefore have a different aim than
out-of-band signal injection attacks.

The term {\em (sensor) spoofing}~\cite{petit_remote_2015, park_pump_2016,
yan_autonomous_vehicles_2016, tu_injected_actuation_2018} was also avoided for
similar reasons: it has an overloaded
meaning in authentication contexts and with in-band signal injection
attacks~\cite{shoukry_non_invasive_2013, davidson_controlling_2016}.
Moreover, it does not capture the physical aspect of
injections, and does not accurately describe coarse-grained attacks which lead
to saturation of a sensor.

Other terms used have been
specific to the particular channel which is being exploited, including
{\em induction attacks}~\cite{selvaraj_induction_embedded_2018},
{\em acoustic resonance}~\cite{shahrad_acoustic_2018},
and {\em (acoustic) non-linearity}~\cite{zhang_dolphin_2017, roy_backdoor_2017,
roy_inaudible_2018}. Such terms were not selected because they are channel-specific,
and focus on the mechanism of the attack, rather than the effect.
Similarly, methodology-inspired terms which have been avoided include
Radio Frequency Injection (RFI), Direct Power Injection (DPI), and other
terminology that arises in immunity or susceptibility
literature against (non-adversarial) electromagnetic
interference (EMI)~\cite{sutu_statistics_demodulation_rfi_1983, ghadamabadi_demodulation_rfi_1990,
boyer_modeling_power_injection_2007, boyer_modelling_2007, gago_emi_susceptibility_2007,
gao_improved_dpi_2011, ayed_failure_mechanism_adc_2015, ayed_immunity_modeling_adc_2015,
boyer_modeling_2016, kennedy_flash_adc_emi_2018, pouant_modeling_2018}.

Finally, the term {\em transduction attacks},
proposed by Fu and Xu~\cite{fu_trusting_sensors_2018} to mean attacks which ``exploit
a vulnerability in the physics of a sensor to manipulate its output or induce
intentional errors'' has not yet received mainstream recognition. It also does not
necessarily make it clear that the attack may target the interface
between the sensor and the rest of the system, instead of just the sensor itself.
The various terms which have been used to describe out-of-band signal injection
attacks are shown in Table~\ref{table:terminology}, along with example references.

Since the majority of attacks in the literature are
on sensors rather than actuators, we will refer to both of them collectively
as sensors for brevity, and will distinguish between the two only when it is
necessary to do so, i.e., when there is a divergence in the
attack methodology.

\begin{table}[!t]
    \centering
    \caption{Terminology used by different works to describe
             out-of-band signal injection attacks.}
    \begin{tabular}{@{}ll@{}}
    \toprule
    \textbf{Terminology}    & \textbf{Example References}\\
    \midrule
    \textbf{Injection}      & \cite{markettos_trng_injection_2009,
                                    kune_ghost_2013,
                                    kasmi_iemi_2015,
                                    trippel_walnut_2017,
                                    nashimoto_sensor_2018,
                                    osuka_em_2019,
                                    tu_injected_actuation_2018,
                                    giechaskiel_framework_2019} \\
    \textbf{Intentional Interference}
                            & \cite{bayon_trng_em_2012,
                                    kune_ghost_2013,
                                    kasmi_iemi_2015,
                                    son_rocking_2015,
                                    bayon_fault_2016,
                                    bolton_blue_2018,
                                    esteves_remote_2018,
                                    osuka_em_2019} \\
    \textbf{Non-Linearity} & \cite{zhang_dolphin_2017,
                                    roy_backdoor_2017,
                                    roy_inaudible_2018,
                                    yan_cuba_2019} \\
    \textbf{Spoofing}       & \cite{petit_remote_2015,
                                    park_pump_2016,
                                    yan_autonomous_vehicles_2016,
                                    tu_injected_actuation_2018}  \\
    \textbf{Other (See Text)}
                           & \cite{selvaraj_induction_embedded_2018,
                                   shahrad_acoustic_2018,
                                   fu_trusting_sensors_2018} \\
    \bottomrule
    \end{tabular}
    \label{table:terminology}
\end{table}

\section{System and Adversary Model}
\label{sec:model}

Systems depend on sensors and actuators to interface with their external environment.
They therefore require a conversion of a physical property (e.g., temperature or speed)
to or from an electrical quantity (such as voltage or resistance). This electrical
measurement is typically analog in nature, and is digitized by an Analog-to-Digital
Converter (ADC) before it is processed. Although modern cryptography has
mostly solved the problem of secure communication between digital interfaces,
there is no way to authenticate the measurement itself, or the analog component
of the connection between the sensor or actuator and a microcontroller. This lack of
authentication, coupled with hardware imperfections, can be exploited for
out-of-band signal injection attacks.

Conceptually, the sensor, the ADC,
and the microcontroller perform logically distinct functions, but
all three can be fully encapsulated into the same Integrated Circuit (IC) chip.
Although this chip presents a digital interface to third parties, which can
be protected by cryptographic protocols, the sensor itself
can still be vulnerable to out-of-band signal injection attacks. For example,
acoustic attacks targeting the resonant frequencies of gyroscopes and
accelerometers have proven to be effective even against
digital ICs~\cite{son_rocking_2015, trippel_walnut_2017, tu_injected_actuation_2018}.

As explained in Section~\ref{sec:terminology}, attackers are not allowed to
manipulate the property being measured itself (e.g., the radiation being
measured by a Geiger counter): in the language of Shoukry et
al.~\cite{shoukry_pycra_2015}, the property measured itself
is {\em trusted}, although the measurement itself is not.
Nonetheless, attackers are allowed to transmit signals outside of the limits
being sensed, which are still interpreted as valid measurements. For example,
an attacker can produce ultrasound waves which are picked up by a microphone
recording human speech~\cite{roy_inaudible_2018, zhang_dolphin_2017,
roy_backdoor_2017}, or shine infrared (IR) light into a camera capturing
the visible part of the spectrum~\cite{petit_remote_2015}.
However, attacking a microphone with audible sound, or
a camera with visible light is not allowed under this threat model,
since the attacker is manipulating the property being sensed
in-band. A secondary goal for some attacks is therefore
undetectability or {\em concealment}~\cite{shoukry_pycra_2015}.
Attacks also need to be {\em non-invasive}~\cite{shoukry_pycra_2015}, and preclude direct
physical access to the system under attack. It should be noted that different
attack techniques have different distance requirements, with
electromagnetic attacks theoretically having a longer range compared to
optical and acoustic attacks. We defer the discussion of distance
and related considerations to Section~\ref{sec:future}.

\begin{figure}[t!]
    \centering
    \includegraphics[width=\linewidth]{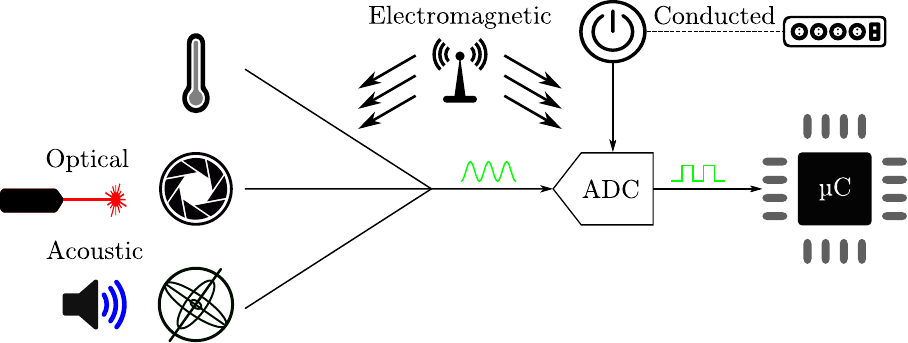}
    \caption{System model for out-of-band signal injection attacks.
             Remote and conducted adversarial electromagnetic emanations,
             optical emissions, and acoustic waves
             can attack the sensors themselves,
             or the interfaces connecting sensors to microcontrollers
             through Analog-to-Digital Converters (ADCs).}
     \label{fig:channels}
\end{figure}

Figure~\ref{fig:channels} summarizes the channels that have successfully
been exploited in the literature thus far.
Some of the out-of-band signal injection attacks use electromagnetic (EM) waves
to penetrate the wires connecting sensors and microcontrollers (Section~\ref{sec:em}),
or the power circuit of the device (Section~\ref{sec:conducted}). The same effects
can sometimes also be achieved through shared power lines (conducted attacks).
Finally, other attacks may target the sensors themselves
through sound (Section~\ref{sec:acoustic}), and alternative
means such as infrared light (Section~\ref{sec:others}).

The effectiveness of different attacks, however, has mostly been evaluated
in a qualitative fashion so far  (e.g., whether the system was tricked
into performing an action or not). Recent research, however, has attempted
to mathematically define security against out-of-band signal injection attacks,
and therefore quantify their success~\cite{giechaskiel_framework_2019}.
Specifically, Giechaskiel et al.~\cite{giechaskiel_framework_2019} introduced probabilistic
definitions which attempt to capture the fidelity with which an adversary can make
a target waveform appear at the input of the microcontroller of Figure~\ref{fig:channels}.
These definitions address both coarse-grained attacks which merely
disrupt sensor measurements and fine-grained attacks with precise waveform
injections. Although the definitions abstract away from specific hardware
considerations, they still model the discrepancy between an adversary's
target waveforms and the actual signal transmitted, which is necessary
due to the behavior of the underlying circuits.

The specific circuit properties that allow adversarial signals to be injected into
a device vary with the injection method and module targeted. For example,
EM attacks typically depend on
unintentional antennas in Printed Circuit Board (PCB) traces~\cite{kune_ghost_2013},
while acoustic attacks exploit resonance in gyroscopes~\cite{son_rocking_2015} and
accelerometers~\cite{trippel_walnut_2017}. As each attack exploits unique
properties of the target device, the details of specific hardware
imperfections in sensors and other modules are only expanded upon in subsequent
sections. This section instead discusses common features of different attacks
at a high level. One of these commonalities is that the vulnerability of systems
to out-of-band signal injection attacks depends on both: (a) how adversarial signals
are received by the devices under attack; and (b) how these signals are digitized.

\begin{figure}[t!]
    \centering
    \includegraphics[width=\linewidth]{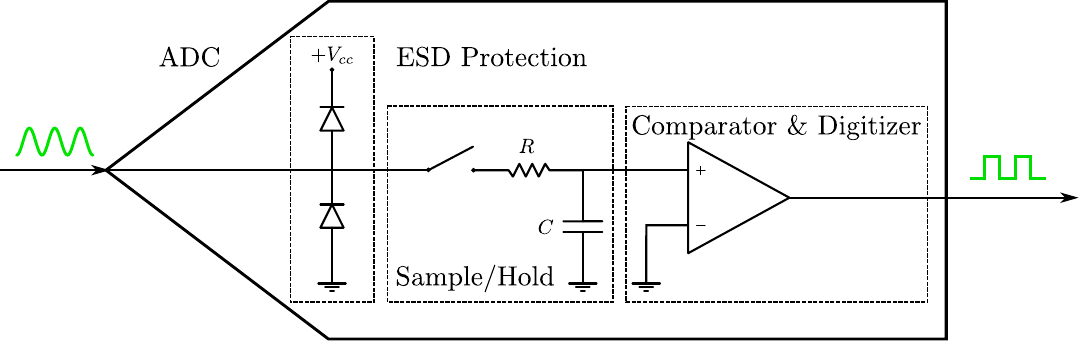}
    \caption{Analog-to-Digital Converter (ADC) model: non-linearities due to
             Electrostatic Discharge (ESD) protection diodes and amplifiers (e.g.,
             comparators) counteract low-pass filtering effects of the sample-and-hold
             mechanism and can unintentionally demodulate high-frequency
             input signals.}
     \label{fig:adc}
\end{figure}

Giechaskiel et al.~\cite{giechaskiel_framework_2019} recently proposed a general circuit
model which uses two transfer functions to separate these two aspects of
vulnerability. The first transfer function describes circuit-specific
transformations that an adversarial signal undergoes. For example, for EM attacks,
these transformations include the (unintentional) low-power, low-gain antenna-like behavior
of PCB traces connecting sensors to ADCs~\cite{kune_ghost_2013, giechaskiel_framework_2019}.
The second transfer function, on the other hand, is ADC-specific, and
summarizes the artifacts of the digitization process. These two transfer functions
dictate that for a successful injection, attacker signals typically
need to be transmitted over high-frequency
carriers, and be demodulated into low-frequency, meaningful waveforms.
According to the work of Giechaskiel et al., components within ADCs are the
culprits for these demodulation effects~\cite{giechaskiel_framework_2019}.
The main constituents of an ADC that contribute to its demodulation
characteristics are therefore summarized in Figure~\ref{fig:adc}.

According to Pelgrom, an ADC uses three basic components to convert analog
signals into digital ones: a ``sample- or track-and-hold circuit where the
sampling takes place, the digital-to-analog converter and a
level-comparison mechanism''~\cite{pelgrom_adc_book_2017}.
Level-comparison amplifiers contribute to
the demodulation properties of ADCs~\cite{giechaskiel_framework_2019} due to
non-linear distortions, including {\em harmonics} and
{\em intermodulation products}~\cite{redoute_emc_2009}.

Harmonics are responsible for producing ``spectral components at  multiples
of the fundamental [input] frequency''~\cite{redoute_emc_2009}. As an example,
for a sinusoidal of angular frequency $\omega=2\pi f$, harmonics transform
the input $v_{in}=\hat{v}\cdot \sin(\omega t)$ into:
\begin{align}
  v_{out} &= \left(\frac{a_2\hat{v}^2}{2}+\frac{3a_4\hat{v}^4}{8}+\cdots\right)
           +\left(a_1\hat{v}+\frac{3a_3\hat{v}^3}{4}+\cdots\right)\sin(\omega t) \nonumber \\
          &- \left(\frac{a_2\hat{v}^2}{2}+\frac{a_4\hat{v}^4}{2}+\cdots\right)\cos(2\omega t)
           + \cdots \label{eq:harmonics}
\end{align}
As Equation~\eqref{eq:harmonics} shows, the output also contains a
Direct Current (DC) component, which depends solely on
the ``even-order nonlinear behavior''~\cite{redoute_emc_2009} of the system.

DC shifts can also be the result of reverse-biased diodes at the input of an ADC.
These diodes protect the circuit from Electrostatic Discharge (ESD)
by clamping negative voltages to ground, and inputs exceeding the
maximum allowed voltage to $V_{cc}$. This behavior of ESD diodes
can cause a (non-linear) DC shift~\cite{redoute_emc_2009}, which
attackers can also exploit~\cite{selvaraj_induction_embedded_2018}.

Intermodulation distortions, on the other hand, arise when the input signal contains signals
of two different frequencies. For example, a sum of two sinusoidals
$v_{in}=\hat{v}_1\cdot \sin(\omega_1 t)+\hat{v}_2\cdot \sin(\omega_2 t)$
may represent an adversarial signal laid on top of the legitimate
sensor signal~\cite{giechaskiel_framework_2019}. Non-linearities and trigonometric
identities would then dictate that the output signal contains
frequencies of the form $n\omega_1 \pm m\omega_2$ for integers $n,m$.
As Giechaskiel et al.\ note, both types of ``non-linearities demodulate attacker waveforms,
even when they are modulated on high-frequency carriers''~\cite{giechaskiel_framework_2019}.
This fact makes them crucial for out-of-band signal injection attacks.

The final component of ADCs which is relevant to the adversarial injections
studied in this article is the sample-and-hold circuitry shown in Figure~\ref{fig:adc}.
In its simplest form, the sample-and-hold mechanism consists of a resistor and
a capacitor ({\em $RC$ circuit}) connected to the input of the ADC. The transfer
function of the voltage across the capacitor is therefore $H_{S/H}(j\omega)=\frac{1}{1+j\omega RC}$.
This dictates that as the angular frequency $\omega$ increases, the
gain $G_{S/H}=\frac{1}{\sqrt{1+\left(\omega R C\right)^2}}$ is
reduced~\cite{giechaskiel_framework_2019}. To put it differently,
the sample-and-hold mechanism acts as a low-pass filter,
counteracting the {\em aliasing} effect,
which occurs when input signals are faster than half the sampling rate
of the ADC ({\em Nyquist frequency}). One would expect this filtering behavior to
reduce the vulnerability of systems to out-of-band signal injection
attacks. However, the filter's cutoff frequency in practice ``is often
much higher than the sampling rate of the ADC''~\cite{giechaskiel_framework_2019},
necessitating additional anti-aliasing filters before the ADC input~\cite{pelgrom_adc_book_2017}.

Overall, imperfections in the sensors themselves, the ADCs,
or the connections between them can result in high-frequency signals being
interpreted as meaningful low-frequency ones. As modulation over
high-frequency signals is often necessary to enter the
targeted circuit~\cite{kune_ghost_2013, giechaskiel_framework_2019},
the demodulation properties of ADCs allow remote attackers without
physical access to inject signals into a system in an out-of-band fashion.
The subsequent sections discuss these imperfections in greater detail, with a
focus on the specific method of injection.

\section{Electromagnetic Transmissions}
\label{sec:em}

The antenna-like behavior of wires and traces is extensively studied in the
fields of Electromagnetic Compatibility (EMC) and Electromagnetic Immunity.
This is done to ensure interoperability between the various household electrical appliances, by
guaranteeing that devices neither cause nor are susceptible to undue
interference~\cite{montrose_emc_1999}. Such research has shown an inverse
relationship between the length of microstrip PCB traces and the frequencies
at which the traces are resonant~\cite{leone_pcb_emi_1999}. However, to
better predict the response of PCB traces to external EM fields,
many additional parameters are important in practice. For instance,
``field incidence, polarization angles, and the magnitude and phase of the
impedances loading the microstrip
terminations''~\cite{lagos_worst_i_2011} are all useful in modeling trace behavior.

Although going into the details of such antenna models is outside the scope of this article,
the effects of unintentional antennas have been central in the security community.
Until now, unintentional {\em transmitting} antennas have been key
for side-channel analysis: data-dependent emissions can reveal the information
processed by a device to a remote attacker~\cite{kocher_introduction_2011}.
However, wires conversely acting as unintentional {\em receiving}
antennas have only become a focal point of research more recently: out-of-band
signal injection attacks have demonstrated that the antenna-like behavior of wires
between sensors and microcontrollers can result in adversarial EM signals
being interpreted as legitimate measurements.

As the systems targeted are not
intended for communication, the phenomenon is known as {\em back-door coupling}
\cite{backstrom_susceptibility_2004, giri_classification_2004,
mansson_vulnerability_2008, kune_ghost_2013, wolfgramm_field_2015}. In
back-door coupling, the ``radiation couples through imperfections (apertures) in an
electromagnetic shield, giving rise to a diffuse and complex field pattern
within the shielded structure''~\cite{backstrom_susceptibility_2004}.
Consequently, predicting the susceptibility of systems against back-door coupling is a hard
task ``without detailed testing, although properties averaged over frequency
bands can be predicted''~\cite{benford_hpm_2016}.
In other words, although the resonant behavior of simple geometric structures (e.g., lines
and rectangles) has been extensively studied~\cite{pozar_microwave_2011},
extensive experiments are necessary to identify the extent to which intermodulation
products appear~\cite{wolfgramm_field_2015}.
Such products of diodes and other non-linear components
can act as potential envelope detectors causing systems to take the wrong
safety-critical actions~\cite{kune_ghost_2013}. As a result, these effects are
a concern for more than just compliance with EMC regulations.

Implantable medical devices (IMDs) are an example of a safety-critical system
where external electromagnetic interference (EMI) can cause physical harm
to people. As a result, there is extensive research on the EMI behavior
of various IMDs~\cite{irnich_interference_1996, hayes_interference_1997,
pinski_interference_i_2002, pinski_interference_ii_2002, barbaro_mechanisms_2003,
cheng_effects_2008, cohan_environmental_2008,
lee_clinically_2009, seidman_vitro_2010,
misiri_electromagnetic_i_2012, misiri_electromagnetic_ii_2012,
buczkowski_influence_2013, driessen_electromagnetic_2019}. Some of these works
have even pinpointed the properties of ``non-linear circuit elements''
in pacemakers as the culprits for demodulating RF signals
produced by cell phones~\cite{barbaro_mechanisms_2003, censi_interference_2007}.
However, the consequences of intentional out-of-band electromagnetic signal injection
attacks on IMDs were only first identified in 2009 by Rasmussen et al.\ in the
context of a distance-bounding protocol~\cite{rasmussen_proximity_based_2009}.

The proposed protocol used ultrasound transmissions to place guarantees
on the distance between two communicating parties,
but it was determined that an EM signal could ``induce a current in
the audio receiver circuit just as if the IMD received a sound
signal''~\cite{rasmussen_proximity_based_2009}. This would break protocol properties
which depend on the speed of sound constant: adversarial
transmissions propagating at the speed of light allow an adversary
to operate from a longer distance.
This attack is perhaps the first out-of-band electromagnetic signal injection, since
it utilized EM emanations to attack an ultrasound-based protocol: unintentional
antennas were found on the path ``from the reception circuit to the piezo
element'', which was effectively ``working as a
microphone''~\cite{rasmussen_proximity_based_2009}.

Although the attack by Rasmussen et al.\ was more of a side-note to an
otherwise-secure protocol~\cite{rasmussen_proximity_based_2009},
Foo Kune et al.'s seminal 2013 ``Ghost Talk'' paper~\cite{kune_ghost_2013} made
such adversarial injections the focal point of research. It showed that
EM emissions could affect Electrocardiogram (ECG)
measurements and cause IMDs to deliver fatal defibrillation shocks~\cite{kune_ghost_2013}.
Foo Kune et al.\ succeeded in injecting arbitrary analog measurements, making
a marked improvement in the literature compared to coarse replay and jamming attacks on
IMDs~\cite{halperin_pacemakers_2008, gollakota_they_2011, li_hijacking_2011, rushanan_sok_2014}.
Their approach also significantly differed from high-power Intentional Electromagnetic
Interference (IEMI) leading to the transient upset or destruction of commercial
equipment~\cite{backstrom_susceptibility_2004, giri_classification_2004, hoad_trends_2004,
radasky_introduction_2004, delsing_susceptibility_2006, mansson_susceptibility_2008,
mansson_vulnerability_2008, sabath_classification_2008,
brauer_susceptibility_2009, palisek_high_2011}.

The attack on IMDs by Foo Kune et al.~\cite{kune_ghost_2013} used low-power
($\le\SI{10}{\watt}$), low-frequency
($\si{\kilo\hertz}$ range) EM signals which coupled to the leads of ECGs and
Cardiac Implantable Electrical Devices (CIEDs). In the open air, the distance
achieved was up to $\SI{1.67}{\meter}$, but when submerging the devices in
saline (to approximate the composition of the human body), successful attacks were limited to
less than $\SI{10}{\centi\meter}$. The signals emitted targeted the baseband
(i.e., the frequency of operation) of the IMDs directly, and thus did not
make use of the non-linearities identified above and in Section~\ref{sec:model}.
However, this attack should still be considered out-of-band, as the IMD leads
were meant to require physical contact for measurements,
and should not be reacting to remote electromagnetic transmissions.
In other words, this attack is more akin to coupling to a multimeter's probes
to cause wrong voltage readings remotely, rather than causing interference to
a WiFi device by transmitting at $\SI{2.4}{\giga\hertz}$: although both
require external stimuli, the mode of injection is different from the
intended mode of operation.

Foo Kune et al.\ also conducted out-of-band attacks against webcams and bluetooth headsets
that were up to $\SI{1}{\meter}$ away from an $\SI{80}{\milli\watt}$
source~\cite{kune_ghost_2013}. The authors
transmitted modulated signals over high-frequency carriers (in the hundreds
of $\si{\mega\hertz}$) that make use of unintentional
antennas on the path between the microphone and the amplifier. Non-linearities
then demodulated the input signals and produced intelligible audio output.
This output overpowered legitimate conversations via the headset and
fooled music identification services as well as automated dial-in services (by emulating key presses via
modulated Dual Tone Multiple Frequency (DTMF) signals)~\cite{kune_ghost_2013}.

\begin{figure}[t!]
    \centering
    \includegraphics[width=\linewidth]{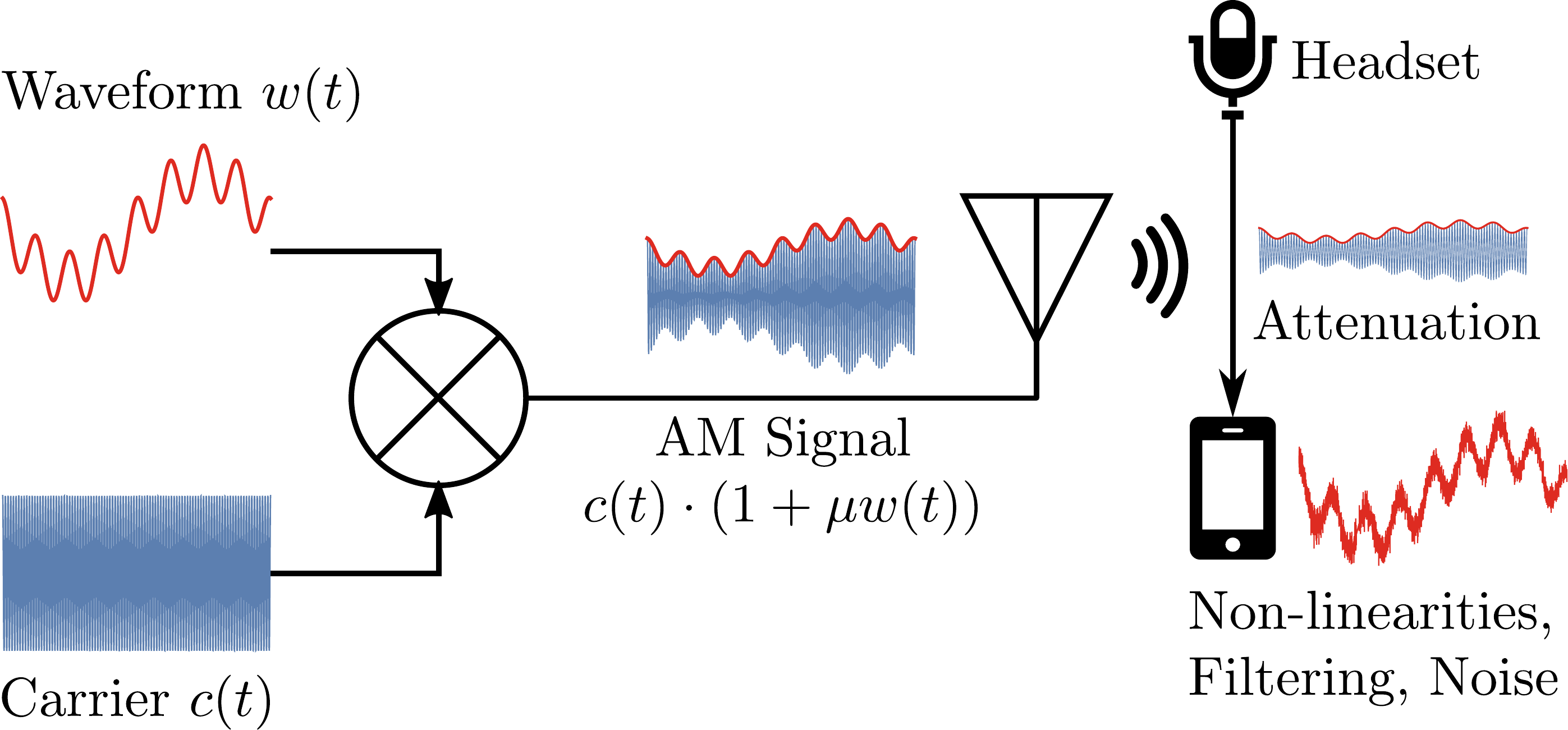}
    \caption{Basic operating principle of an electromagnetic out-of-band
             signal injection attack against a microphone. Amplitude-Modulated
             (AM) signals are transmitted
             using an antenna, and are picked up (and attenuated) by headphone cables or
             PCB traces. Non-linearities in amplifiers coupled with low-pass
             filters remove the carrier wave and down-convert the
             target signal. The demodulated signal can be used to
             break ultrasound protocols~\cite{rasmussen_proximity_based_2009},
             fool music services~\cite{kune_ghost_2013}, or inject
             voice commands~\cite{kasmi_iemi_2015}, despite the
             additional noise introduced.}
     \label{fig:em}
\end{figure}

Kasmi and Lopes Esteves similarly targeted smartphone microphones, but with
a goal of triggering voice commands (e.g., ``OK Google'', ``Hey Siri'')
by emitting Amplitude-Modulated (AM) signals~\cite{kasmi_iemi_2015}. These signals
get picked up by the user's hands-free headset, are then demodulated due to
non-linearities, and finally get executed by the software voice-processing service.
It is interesting to note that by default, ``a long hardware button press
is required for launching the service''~\cite{kasmi_iemi_2015}. However,
a Frequency-Modulated (FM) signal at the same frequency can also
emulate this headphone button press~\cite{kasmi_iemi_2015}, allowing the
attack to be fully carried out remotely.

The attack by Kasmi and Lopes Esteves used {\em front-door coupling}, as the ``radiation
couples to equipment intended to communicate or interact with the external
environment''~\cite{backstrom_susceptibility_2004}. This is because headphones
can be used as FM antennas and can thus not be
effectively shielded. It should be noted that the field strength required
was in the order of $\SIrange{25}{30}{\volt\per\meter}$, which is close to the
limit for human safety, and an order of magnitude higher than the required
immunity level ($\SI{3}{\volt\per\meter}$)~\cite{kasmi_iemi_2015}.
This illustrates that high powers might
still be required for reasonable attack distances:
in a subsequent work, the authors noted that their attack requires a power
of $\SI{40}{\watt}$ for a distance of $\SI{2}{\meter}$, and
$\SI{200}{\watt}$ for a distance of $\SI{4}{\meter}$~\cite{esteves_remote_2018}.

Figure~\ref{fig:em} shows an example of an electromagnetic out-of-band
signal injection attack, which summarizes the attacks against
microphone sensors. It shows that the desired attacker waveform $w(t)$ needs to be
modulated over a high-frequency carrier $c(t)$, so that the signal can be
picked up with relatively low attenuation by the victim device's wires.
For example, amplitude modulation with a modulation depth $0<\mu\le100\%$
can be used to couple to wired
headphones. Through non-linearities in the phone's internal amplifier, as well
as low-pass filtering effects, the target waveform is preserved at the
output of the digital signal processing (DSP) chip. Although this process
introduces noise, the demodulated signal can still be distinguished
by online music services~\cite{kune_ghost_2013}, imitate voice-initiated
commands, which are then executed by the phone~\cite{kasmi_iemi_2015},
or break protocol guarantees~\cite{rasmussen_proximity_based_2009}.
Although the above works targeted microphones, out-of-band
electromagnetic signal injections are not sensor- or device-specific.
For example, subsequent sections discuss proof-of-concept EM-based injections
against temperature sensors~\cite{tu_trick_2019, zhang_detection_2020}.
However, most attacks have primarily used amplitude modulation, leading
to a question that has yet not been addressed:

\begin{open}
What is the optimal modulation scheme for out-of-band electromagnetic
signal injection attacks? In other words, can Frequency Modulation (FM),
Phase Modulation ($\phi$M), or other schemes be used instead of
Amplitude Modulation (AM)?
\end{open}

Another largely-unexplored research area is that of magnetic emissions.
Specifically, magnetic attacks had largely been ignored in the literature,
until Shoukry et al.\ demonstrated that it is possible to confuse
Anti-Lock Braking Systems (ABS)~\cite{shoukry_non_invasive_2013}. This
is done by exposing the magnetic-based wheel speed sensors to an in-band
attacker-generated magnetic field at close proximity~\cite{shoukry_non_invasive_2013}.
Doing so can alter speed measurements, potentially
veering cars off the road~\cite{shoukry_non_invasive_2013}.

Although the work by Shoukry et al.\ is in-band and limited in
distance, it inspired subsequent work in out-of-band attacks:
Selvaraj et al.\ recently conducted the first attacks on actuators
(rather than sensors) through EM transmissions~\cite{selvaraj_induction_embedded_2018}.
Specifically, an unmodulated sawtooth waveform was chosen to cause a
``sharp decrease, for a very small amount of time'' at the target
servo~\cite{selvaraj_induction_embedded_2018}. Because the servo
is controlled using Pulse Width Modulation (PWM), a waveform of the
same frequency ($\SI{50}{\hertz}$) therefore results in a one-way (clockwise)
rotation.

This attack has a few limitations: changing the attacking frequency
to $\SI{60}{\hertz}$ causes the servo to ``change positions
randomly''~\cite{selvaraj_induction_embedded_2018}. Moreover, relatively
high powers are required: a $\SI{10}{\volt}$ (peak-to-peak) waveform
is insufficient, so a $\SI{50}{\watt}$ amplifier and a 1-to-6 step-up
transformer are necessary. Moreover, one of the servo wires is wrapped
around the toroid transferring the EM signal. Although
``the same effect was observed when a length of the wire was placed
within a solenoid'', ``producing an effect at a distance requires the
proper selection of a field directivity element''~\cite{selvaraj_induction_embedded_2018}:

\begin{open}
How can one precisely control an actuator in both directions, and at a distance?
\end{open}

Selvaraj et al.\ additionally proposed an analytical model of
electromagnetic induction attacks for sensors and actuators, with a focus on
the magnetic rather than the electric field~\cite{selvaraj_induction_embedded_2018}.
To support their model, they further conducted
experiments against General Purpose Input/Output (GPIO)
pins of microcontrollers in analog and digital modes. They showed that
$\SI{1.82}{\watt}$ transmissions of unmodulated signals
at frequencies between $\SIrange{0}{1000}{\mega\hertz}$ can result in a DC
offset, even when the microcontroller is at a distance of up to $\SI{1}{\meter}$
from the source. This indicates that an adversary can successfully
inject signals over a wide range of frequencies, without having
precisely determined the resonance behavior of the system.

Selvaraj et al.~\cite{selvaraj_induction_embedded_2018} were only concerned with
the average power received and not time-dependent signals. This is in contrast
to work on the demodulating effects of amplifiers (Section~\ref{sec:conducted}),
which depends on inter-modulation products and harmonics.
Instead, ESD diodes were identified as the culprits for the resulting DC offset,
due to clipping non-linearities. However, it is not clear whether the same
methodology can induce attacker-desired, time-varying waveforms through
modulated (in amplitude or otherwise) transmissions: in the
language of Giechaskiel et al.~\cite{giechaskiel_framework_2019},
Selvaraj et al.\ performed an {\em existential injection} which disturbs the
ADC readings, but not a {\em selective injection} of attacker-chosen waveforms,
unlike the earlier work of Foo Kune et al.~\cite{kune_ghost_2013}.

As a final point of note, researchers have identified that coupling into the
wiring interconnects within ICs is possible~\cite{lagos_worst_i_2011}. However,
this disturbance ``can be neglected up to several gigahertz'', since
ICs are ``usually smaller than a few centimeters''~\cite{lagos_worst_i_2011}.
This leads to another research question:

\begin{open}
Is it possible to conduct out-of-band signal injection attacks
into digital ICs which integrate sensors, ADCs, and microcontrollers?
\end{open}

\noindent Although this question has been answered in the affirmative for
acoustic attacks (Section~\ref{sec:acoustic}), it remains open for
electromagnetic ones.

\section{Conducted Signals}
\label{sec:conducted}

A different class of out-of-band signal injection attacks
requires an indirect physical connection between the attacker and
the victim, such as a shared power line. Unlike their radiated counterparts
of Section~\ref{sec:em}, these {\em conducted} attacks do not require signals to be
picked up by unintentional receiving antennas in the path between sensors and
microcontrollers. Instead, signals are propagated along conductors primarily on
the powering circuit, which can transfer through crosstalk or
coupling to paths containing non-linearities. This propagation of electrical
disturbances through structures
and cables is studied in Transmission-Line Theory~\cite{parfenov_conducted_2004,
mansson_propagation_2007, mansson_propagation_2008}, and through
the Baum-Liu-Tesche (BLT) equation~\cite{tesche_example_2005,
tesche_development_2007, guo_analysis_2015}.

Much like electromagnetic attacks, out-of-band conducted signal injection attacks
have also been primarily experimental in nature. Their methodology often
follows that of susceptibility literature, which predicts a device's
response to high-frequency radio signals. Systems tested include microcontrollers,
ADCs, and other embedded devices which contain I/O and power pins.
The goal of such research is to quantify immunity to
radiated and conducted EM disturbances, and is typically
concerned with the average power received by the embedded system,
similar to the work by Selvaraj et al.~\cite{selvaraj_induction_embedded_2018}
summarized in Section~\ref{sec:em}.

To avoid legal and practical considerations related to
electromagnetic transmissions, the experimental approach followed is known as
Direct Power Injection (DPI),
and consists of injecting harmonic disturbances from a few $\si{\kilo\hertz}$
to a couple of $\si{\giga\hertz}$ and measuring the relationship between forward power
and frequency. Multiple works have shown that as the frequency of the input
increases, immunity to DPI also increases~\cite{boyer_modeling_power_injection_2007,
lafon_immunity_modeling_2010, gao_improved_dpi_2011,
ayed_failure_mechanism_adc_2015, ayed_immunity_modeling_adc_2015,
kennedy_flash_adc_emi_2018}.\footnote{Immunity behavior is different for EMI-induced offsets
through the ground plane for amplifiers~\cite{richelli_susceptibility_2016}
and precision voltage references~\cite{richelli_measurements_2017}.
See the survey by Ramdani et al.~\cite{ramdani_electromagnetic_2009}
for more information on RF immunity models.}
In other words, higher frequencies generally require higher forward power
injections for the same level of susceptibility. This was also true
of the (remote) injections by Selvaraj et al.~\cite{selvaraj_induction_embedded_2018},
discussed above.

A similar methodology can be applied to evaluate the demodulation characteristics
of amplifiers and transistors~\cite{sutu_statistics_demodulation_rfi_1983,
ghadamabadi_demodulation_rfi_1990, gago_emi_susceptibility_2007,
pouant_modeling_2018}, and therefore better predict
the fidelity with which attackers can inject target waveforms, both
in the conducted and in the radiated settings. This was recently done by Giechaskiel
et al.~\cite{giechaskiel_framework_2019} for six ADCs, with
a view on how to exploit the demodulating effect for out-of-band signal
injection attacks. It was shown that ADCs of three different types
from four manufactures, and with different resolutions and sampling frequencies
can all demodulate AM waveforms~\cite{giechaskiel_framework_2019}.
Generally, it was determined that the fundamental frequency persists along with its harmonics
and some high-frequency components, even for carriers which
are multiple times the ADCs' sampling and cutoff frequencies~\cite{giechaskiel_framework_2019}.

It is worth noting, however, that the different ADCs do not behave identically.
For instance, some ADCs require fine-tuning of the carrier frequency,
with $\SI{100}{\hertz}$ making a difference as to whether the injected signal
is fully demodulated or not~\cite{giechaskiel_framework_2019}.
On the other hand, some ADCs are vulnerable across the spectrum, i.e., for all frequencies
which do not get severely attenuated to filtering effects~\cite{giechaskiel_framework_2019}.
Moreover, as discussed in an extended version of the same
paper~\cite{giechaskiel_framework_arxiv_2019}, some ADCs ``result in more sawtooth-like
output'', and are therefore ``more resilient to clean sinusoidal injections''.
Finally, the same extended version demonstrates that attacks can also be
performed remotely, without following DPI methodology: a $\SI{10}{\deci\bel m}$
($\SI{10}{\milli\watt}$) transmission can be demodulated by a receiver amplifier at small
distances ($\SI{5}{\centi\meter}$).

In their ``Trick or Heat'' work, Tu et al.~\cite{tu_trick_2019} also conducted
DPI experiments on operation amplifiers, but
with a view on how to exploit rectification effects for out-of-band signal injection
attacks on temperature sensors. They, too, determined that as the frequency increases,
the magnitude of the AC voltage decreases, while ``EMI signals at specific
frequencies induce a significant DC offset''~\cite{tu_trick_2019}.
Moreover, for a given frequency of injection,
power and the induced DC offset are ``locally proportional'', though the rate
of change ``gradually decreases as the power of injected EMI signals grows''~\cite{tu_trick_2019}.
However, power and DC offset are not always positively correlated,
even for remote transmissions: for some frequencies,
the induced DC offset is negative~\cite{tu_trick_2019}.

Having characterized the behavior of individual amplifiers, Tu et al.\
turned their attention to different types of thermal sensors, including
Negative Temperature Coefficient (NTC) thermistors, shielded and unshielded
K-type thermocouples, and Resistance Temperature Detectors (RTDs).
With a $\SI{35}{\deci\bel m}$ ($\SI{3.2}{\watt}$) electromagnetic source, Tu et al.\
succeeded in changing the reported temperature of various devices by at least
$\SI{0.5}{\celsius}$~\cite{tu_trick_2019}. The systems attacked included, among others,
newborn incubators, soldering irons, and 3D printers, which were
placed at distances of up to $\SI{6}{\meter}$.

In some experiments, a thick wall was present between the transmitting
device and an infant incubator under attack.
It was shown that, even in this setup, an adversary can increase the measured skin
temperature by $\SI{3.4}{\celsius}$ or decrease it by $\SI{4.5}{\celsius}$~\cite{tu_trick_2019},
again demonstrating the potentially fatal consequences of out-of-band attacks.
Most of the attacks by Tu et al.\ used unmodulated transmissions,
and therefore only looked at the relationship between frequency and DC offset,
or the relationship between power and DC offset. However, when investigated jointly,
amplitude modulation was capable of causing
selective injections~\cite{giechaskiel_framework_2019}. In other words,
Tu et al.~\cite{tu_trick_2019} spelled ``HI'' in the output of the temperature
sensor by appropriately modulating the amplitude of the transmission.

In a different strand of research, Lopes Esteves and Kasmi demonstrated how to
inject voice commands (``OK Google'') into a smartphone through conducted
means~\cite{esteves_remote_2018}. Specifically, the attack exploited the fact that
on the device's circuit board, the phone's USB charging port is physically
close to the audio frontend, where demodulation (envelope detection) can take
place due to non-linearities~\cite{esteves_remote_2018}.
As a result, back-door coupling occurs, either
due to ``a re-radiation of the interference from the USB circuitry bypassing
the physical isolation by parasitic coupling (crosstalk) or the possible sharing
of the $V_{cc}$ and GND networks on the PCB''~\cite{esteves_remote_2018}.

\begin{open}
What properties of the power circuit and related layout considerations
make systems vulnerable to conducted out-of-band signal injections?
\end{open}

The methodology used was inspired by experiments on the propagation of
conducted disturbances and on EM injections into power cables. Specifically,
amplitude-modulated signals were injected at various locations of the power network,
i.e., at different plug points on the same strip and on extension cords. The phone was left
charging either on a computer USB port, or through a wall adapter. Experiments were
repeated both with a magnetic injection probe (directly coupling to
cables), and a ``custom coupler made with capacitors, resistors and a high-frequency
transformer''~\cite{esteves_remote_2018}. In all cases, it was determined that
the smartphone can demodulate (and execute) commands carried
on the $\SIrange{200}{250}{\mega\hertz}$ range at distances up to
$\SI{10}{\meter}$, even with only a $\SI{0.5}{\watt}$ source. Such conducted attacks therefore
significantly lower the power requirements and increase the injection distance
compared to the same authors' remote EM attack on smartphones~\cite{kasmi_iemi_2015}.

True Random Number Generators (TRNGs) which are based on Ring Oscillators (ROs)
are also vulnerable to conducted signal injection attacks. ROs
are composed of an odd number of logical NOT gates chained together in a ring
formation, where the output of the last gate is used as the input to the first gate.
The value between any two stages of the RO oscillates between true and
false, thus forming a bi-stable loop. The frequency of oscillation
is influenced by the delay of the logic gates and the delay between the RO's
stages, which are in turn influenced by small variations in the
manufacturing process, as well as voltage, and temperature (PVT)~\cite{hajimiri_ro_jitter_1999}.

As a result, by XORing several ring oscillators together, one can exploit the randomness
of the phase jitter to create a TRNG~\cite{sunar_provably_2007}.
However, due to the frequency dependence on voltage, a suitable signal
can lead to frequency locking of the
oscillators~\cite{adler_study_1946, mesgarzadeh_study_2005}, removing
the differences in the randomness of the jitter.
Markettos and Moore first conducted the attack in practice in 2009 by directly
injecting $\SI{24}{\mega\hertz}$ signals into the power supply of two
ring oscillators composed of discrete logic chips~\cite{markettos_trng_injection_2009}.
Moreover, they succeeded in biasing TRNGs even in secure microcontrollers
and smartcards: a sinusoidal wave of $\SI{1}{\volt}$ peak-to-peak
($\SI{2.5}{\milli\watt}$) at $\SI{24.04}{\mega\hertz}$ was enough to
cause a $\SI{5}{\volt}$ EMV Chip and Pin smartcard to fail statistical
tests of randomness~\cite{markettos_trng_injection_2009}.

\begin{figure}[!t]
    \centering
    \resizebox{\linewidth}{!}{
    \subcaptionbox{Unlocked\label{fig:trng_unlocked}}{%
      \includegraphics[width=.47\linewidth]{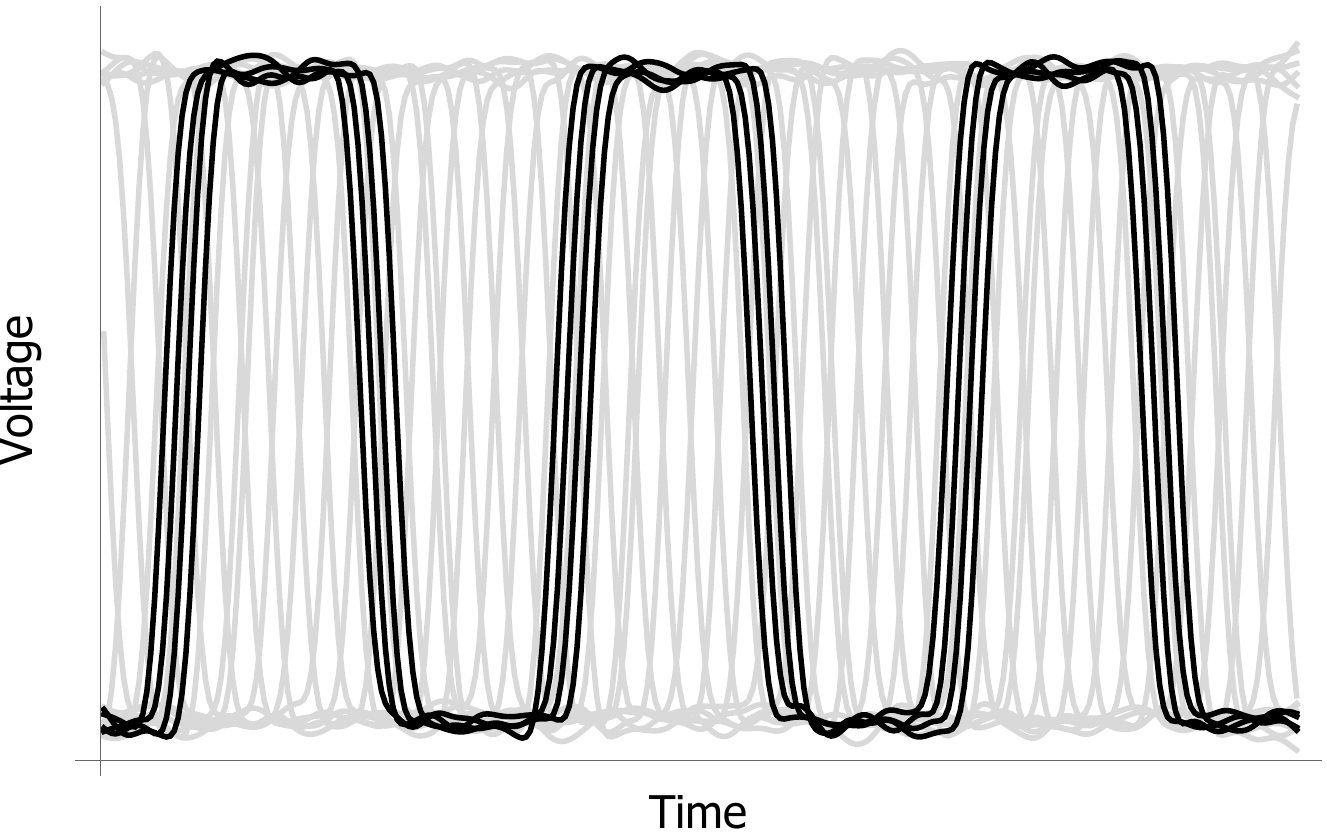}}\quad
    \subcaptionbox{Locked\label{fig:trng_locked}}{%
      \includegraphics[width=.47\linewidth]{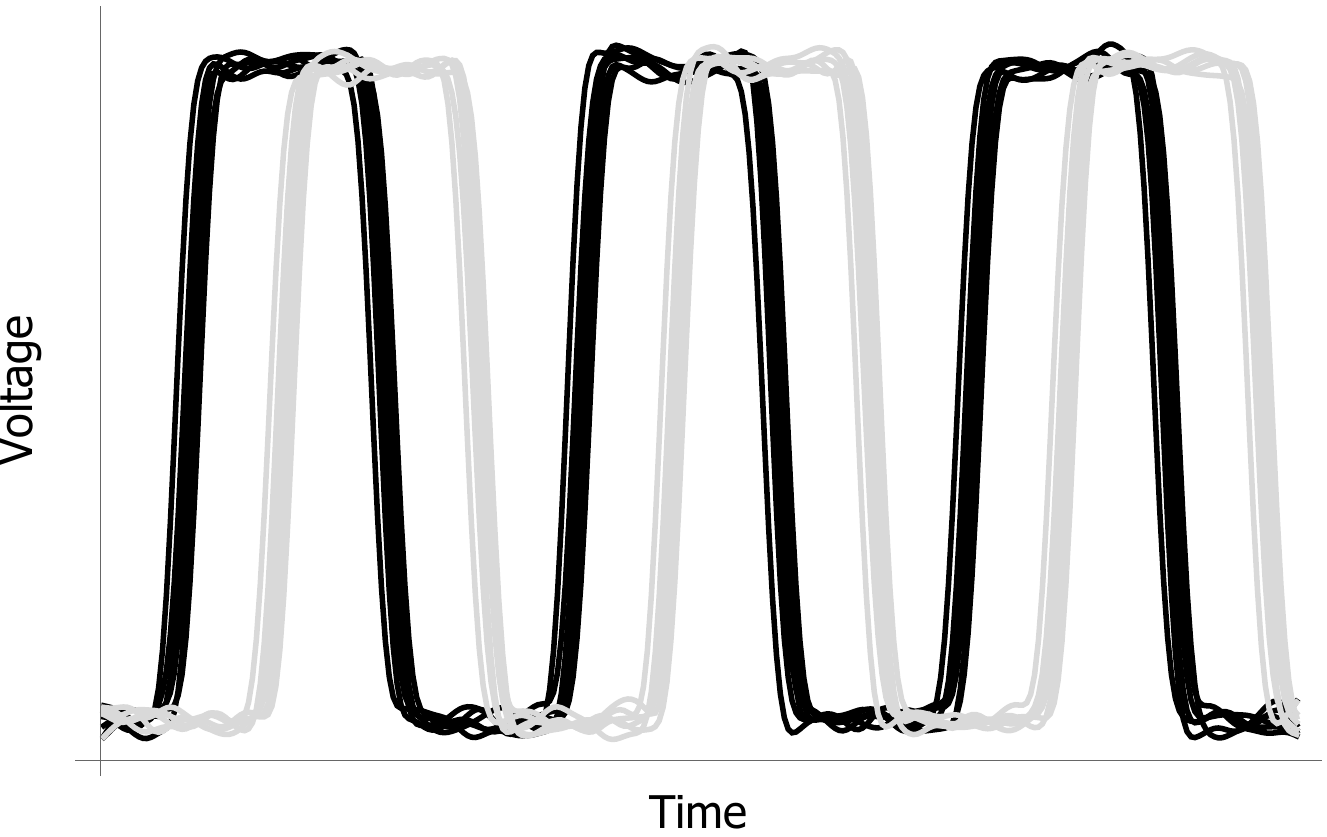}}}
    \caption{Example waveforms for two Ring Oscillators (ROs) with frequency locking
             (\subref{fig:trng_unlocked}) absent or (\subref{fig:trng_locked}) present.}
  \label{fig:trng_state}
\end{figure}

Time-varying signals are not always necessary: a constant (DC) power supply
voltage can also lead to locking of ring oscillators in an under-volted
Field-Programmable Gate Array (FPGA). This is because
there is a ``dependence of the frequency of one oscillator on the current
peaks caused by rising and falling edges of the second oscillator''~\cite{bochard_true_2010}.
Figure~\ref{fig:trng_state} illustrates what happens when two ring oscillators
frequencies lock: during normal operation (Figure~\ref{fig:trng_unlocked}),
the ring oscillator values ``slide past each other, minimising the
likelihood of two rings transitioning together''~\cite{markettos_trng_injection_2009}.
However, when a frequency- or voltage-based attack causes the ROs to lock
(Figure~\ref{fig:trng_locked}), their relationship becomes predictable,
biasing the TRNG.

It should be noted that although under-/over-power attacks are usually considered
fault attacks, in this case the ring oscillators are still functioning properly,
but the entropy of the TRNG is reduced due to less jitter
present~\cite{martin_fault_2015, cao_exploring_2016}. An interesting new class
of such remote under-voltage attacks on TRNGs has recently surfaced.
Because ring oscillators have the potential to increase the delay of FPGA
elements by causing voltage drops, they can also cause timing violations,
thereby reducing the randomness of TRNGs~\cite{mahmoud_timing_2019}. Such an attack
does not require equipment for physical injections. Instead, the adversary only needs
co-located (but logically and physically isolated) circuits on the same FPGA
as the target TRNG. This setup reflects multi-tenant cloud
designs~\cite{mahmoud_timing_2019}, and presents new challenges
for the protection of shared FPGAs against software-only attacks without
physical access.

Although the above attacks generally alter the power supply directly,
the same outcome can be achieved through EM emanations targeting the wires
connecting the various stages of the ring oscillators~\cite{bayon_trng_em_2012,
buchovecka_frequency_2013, bayon_fault_2016}. This requires
micro-probes at very close proximity to the ring oscillators (in the order of
$\SI{100}{\micro\meter}$ from the FPGA packaging), so as to localize the
effects of the injection~\cite{bayon_trng_em_2012, bayon_fault_2016}.
However, TRNGs are also vulnerable against EM injections into power supply cables:
Osuka et al.\ demonstrated that an injection probe
wrapped around the DC power supply cable of a TRNG can also
bias the TRNG~\cite{osuka_em_2019}. Although the design only used two ring
oscillators composed of discrete logic chips,\footnote{
    Much like the original work by Markettos and Moore~\cite{markettos_trng_injection_2009}.
    However, Bayon et al.~\cite{bayon_trng_em_2012, bayon_fault_2016}
    targeted a more realistic TRNG composed of 50 ROs.
} injections were successful even when the probe was placed at a distance
of $\SI{40}{\centi\meter}$ from the ROs, with a power of only
$\SI{25.2}{\deci\bel m}$ ($\SI{0.33}{\watt}$).

\begin{figure}[t!]
    \centering
    \includegraphics[width=\linewidth]{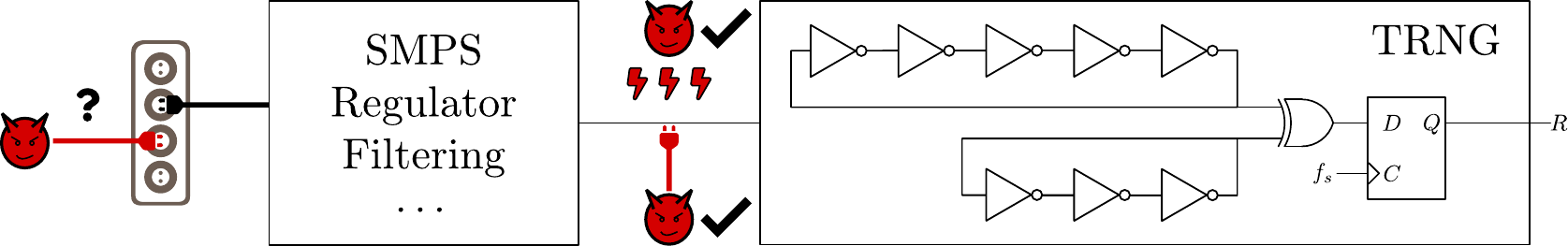}
    \caption{True Random Number Generators (TRNGs) based on Ring Oscillators (ROs)
             are vulnerable to frequency locking: electromagnetic and conducted
             signals into power supply cables can bias the randomness outputs.
             Are attacks through a shared mains power supply network also possible?}
     \label{fig:trng}
\end{figure}

It is worth highlighting that although the conducted voice command injection attack by
Lopes Esteves and Kasmi was performed over shared power lines~\cite{esteves_remote_2018},
all existing attacks on TRNGs bypass AC-to-DC rectification and voltage regulation.
This leads to the following question for future research:

\begin{open}
Is it possible to bias True Random Number Generators through conducted
out-of-band signal injection attacks on the primary side of power supplies
(mains voltage), as shown in Figure~\ref{fig:trng}?
\end{open}

\section{Acoustic Emanations}
\label{sec:acoustic}

\begin{figure*}[!t]
    \centering
    \resizebox{\linewidth}{!}{
    \subcaptionbox{Denial-of-Service\label{fig:gyro_dos}}{%
      \includegraphics[width=.33\linewidth]{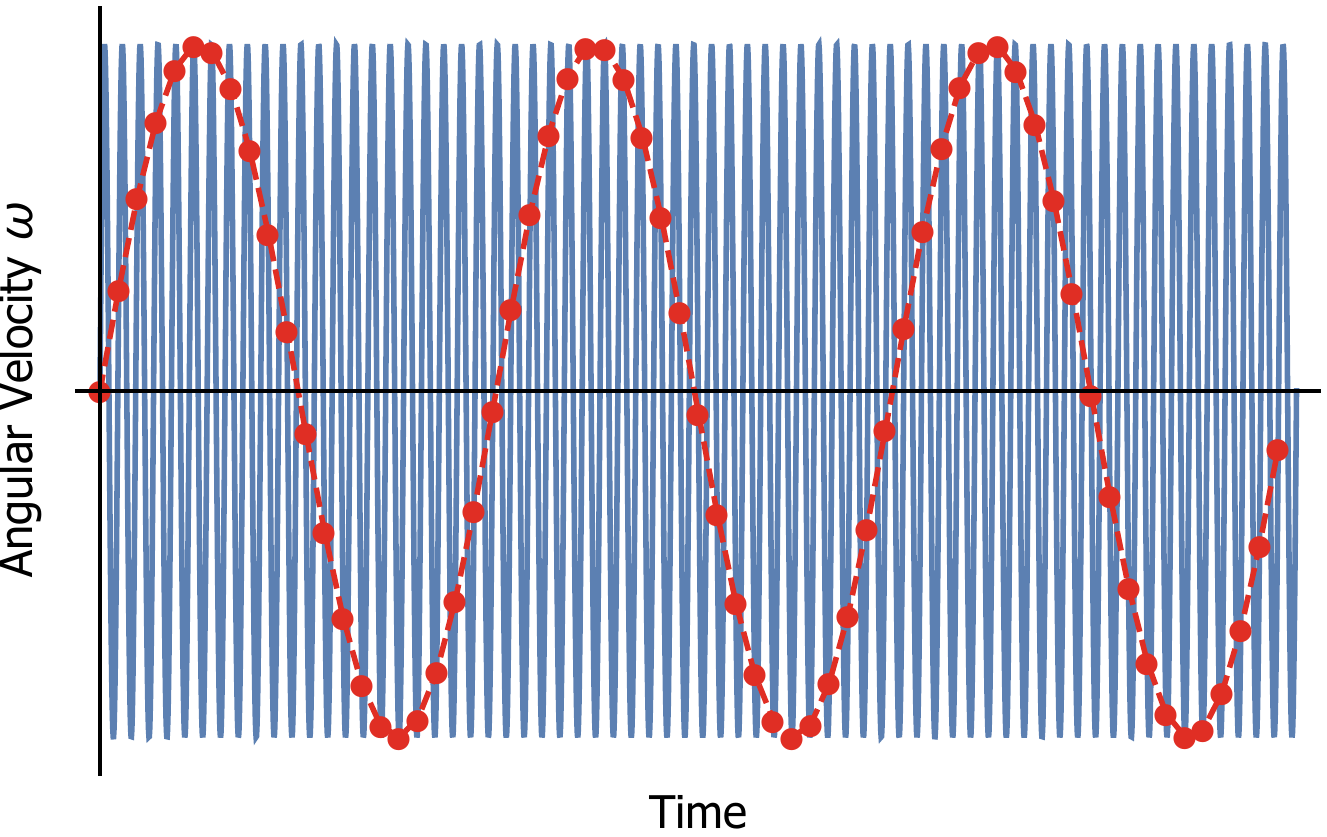}}\quad
    \subcaptionbox{Side-Swing\label{fig:gyro_swing}}{%
      \includegraphics[width=.33\linewidth]{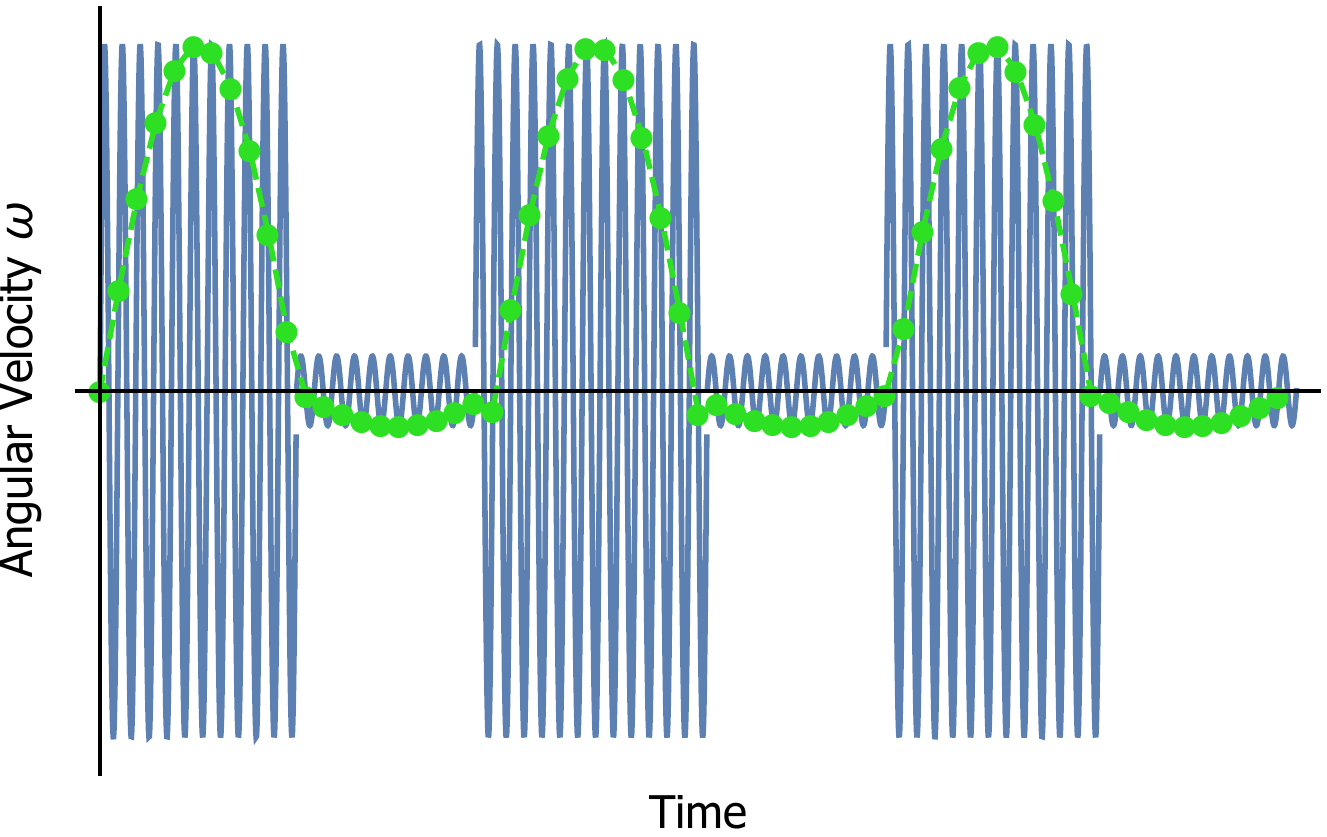}}\quad
    \subcaptionbox{Switching\label{fig:gyro_switch}}{%
      \includegraphics[width=.33\linewidth]{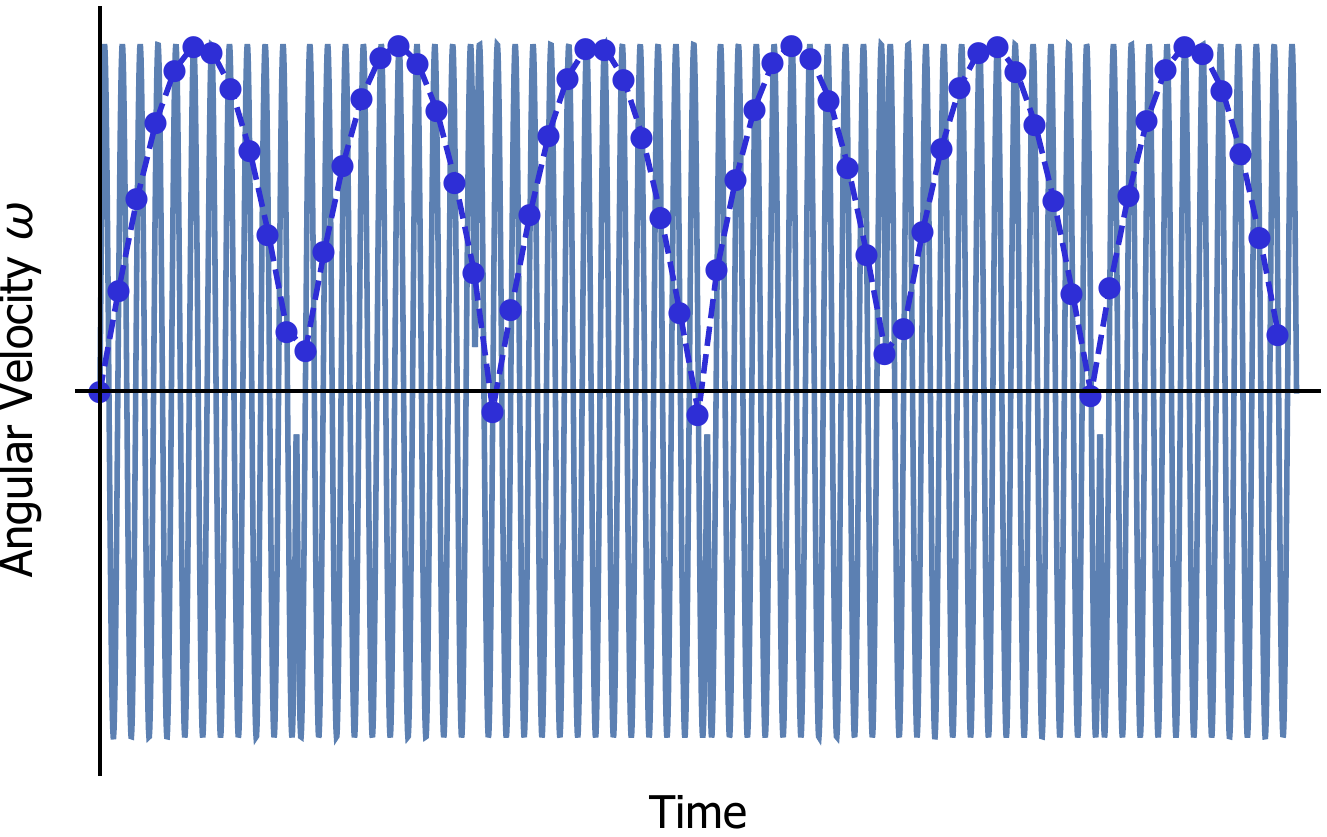}}}
    \caption{Different acoustic injection approaches against gyroscopes.
             For (\subref{fig:gyro_dos}) a Denial-of-Service (DoS) attack,
             a single-tone transmission at the gyroscope's resonant frequency
             suffices. This results in oscillating digital measurements of
             angular velocity, and
             can destabilize equipment~\cite{son_rocking_2015, tu_injected_actuation_2018}.
             To remove the negative measurement components, one can either
             (\subref{fig:gyro_swing}) decrease the transmission amplitude in a {\em Side-Swing}
             attack~\cite{tu_injected_actuation_2018}, or (\subref{fig:gyro_switch}) change
             the frequency of transmission for a {\em
             Switching} attack~\cite{tu_injected_actuation_2018}. The cumulative
             effects of these approaches are shown in Figure~\ref{fig:gyro_sum}.}
  \label{fig:gyro}
\end{figure*}

\begin{figure}[t!]
    \centering
    \includegraphics[width=\linewidth]{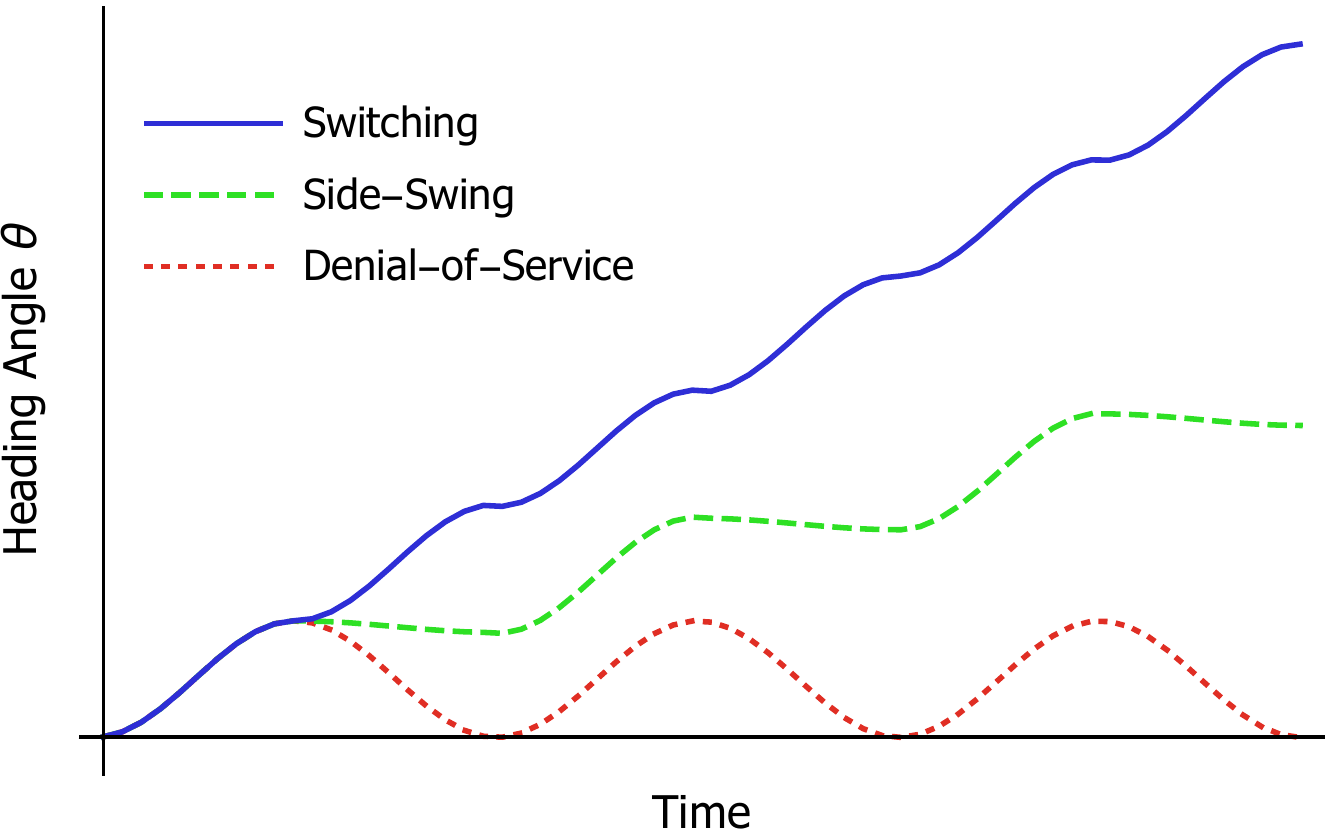}
    \caption{Effects of the three attacks of Figure~\ref{fig:gyro}. In a
             Denial-of-Service (DoS) attack, the accumulating heading angle fluctuates,
             while it continues increasing for both attacks proposed by Tu et
             al.~\cite{tu_injected_actuation_2018}. In Switching attacks,
             the angle increases at twice the rate of Side-Swing attacks.}
     \label{fig:gyro_sum}
\end{figure}

Research into out-of-band acoustic signal injection attacks has primarily
focused on: (a) attacking electro-mechanical
devices by causing vibrations at their resonant frequencies; and (b) exploiting
microphone non-linearities for inaudible voice commands.
In the former category, Micro-Electro-Mechanical Systems (MEMS) gyroscopes and
accelerometers have been a popular target for acoustic resonance attacks.

MEMS gyroscopes operate through ``vibrating mechanical elements
to sense rotation''~\cite{armenise_advances_2011}. In other words, MEMS gyroscopes
contain oscillating structures which, when rotated, appear to have a measurable
force (called the Coriolis force) exerted on them~\cite{acar_mems_2009}. These
mechanical resonators ``generate and maintain a constant
linear or angular momentum'', so that ``when the gyroscope
is subjected to an angular rotation, a sinusoidal Coriolis force at the frequency
of drive-mode oscillation is induced in the sense direction''~\cite{acar_mems_2009}.
This force is exerted in a different direction from the moving direction,
and, depending on the type of the gyroscope, the angular rotation can be estimated
through changes in capacitance, piezoresistive effects, etc.~\cite{acar_mems_2009,
armenise_advances_2011}. Out-of-band acoustic signal injections
transmit sounds at the resonant frequencies of MEMS sensors, causing
them to report incorrect values.

Early research into the properties of MEMS gyroscopes had shown
that high-power acoustic noise at or near the resonant frequency
can degrade the performance of the sensor~\cite{castro_influence_2007,
dean_degradation_2007, dean_characterization_2011}. However, the security effects
of intentional sound transmissions were not explored until the 2015
``Rocking Drones'' paper by Son et al.~\cite{son_rocking_2015}.
Initially, the effect was a simple denial-of-service (DoS) attack on
drones. It was caused by the transmission of single-tone sound waves
at the resonant frequency of drones' gyroscopes, so there was no
control over their movements. The distance was also short,
at $\SI{10}{\centi\meter}$ using a speaker producing a Sound
Pressure Level (SPL) of up to $\SI{113}{\deci\bel}$ at the target frequencies.

Proof-of-concept control was then first
demonstrated in a Black Hat conference presentation against gyroscopes in virtual
reality (VR) headsets and self-balancing vehicles~\cite{wang_sonic_2017}.
Tu et al.'s ``Injected and Delivered'' paper later became the first academic work
to control gyroscopes in a much more fine-grained
fashion~\cite{tu_injected_actuation_2018}. This research allowed control
for long periods of time (up to the minute range), and at long distances
(up to $\SI{7.8}{\meter}$ with a maximum SPL of
$\SI{135}{\deci\bel}$)~\cite{tu_injected_actuation_2018}.

Tu et al.\ noticed that single-tone frequencies (like the ones used by
Son et al.~\cite{son_rocking_2015}) result in an oscillating discrete (digitized)
output, which destabilizes equipment. In other words, a simple transmission
at the resonant frequency is a type of DoS attack because the angular velocity
(as measured by the gyroscope) fluctuates between positive and negative values
(Figure~\ref{fig:gyro_dos}).
However, it is possible to remove these negative components by decreasing
the transmission amplitude during the corresponding measurements. This is
called a {\em Side-Swing} attack, which ``proportionally [amplifies] the induced output in
the target direction and attenuate[s] the output in the opposite
direction''~\cite{tu_injected_actuation_2018} (Figure~\ref{fig:gyro_swing}).

Instead of attenuating signals during half of the transmission period, one can also
control ``the induced output by manipulating the phase of the digital signal with
repetitive phase pacing'' in a {\em Switching} attack~\cite{tu_injected_actuation_2018}
(Figure~\ref{fig:gyro_switch}). As this is accomplished in practice by changing
the tonal frequency instead of attenuating the amplitude, a Switching attack
contributes twice as much to the overall change in direction as a Side-Swing
attack. This is shown by looking at the accumulating heading angle in
Figure~\ref{fig:gyro_sum}. It should be noted that
by accounting for drifts in the sampling rate of the ADC (which are amplified
during adversarial injections~\cite{tu_injected_actuation_2018}), both attacks
can control the gyroscopic output for longer periods of time.

In response to the rising interest in acoustic vulnerabilities,
Khazaaleh et al.~\cite{khazaaleh_vulnerability_2019} created a mathematical model to
explain the resonance response of gyroscopes. They showed that ``the misalignment
between the sensing and driving axes of the gyroscope is the main culprit behind
the vulnerability of the gyroscope to ultrasonic attacks''~\cite{khazaaleh_vulnerability_2019}.
More precisely, because ``the sensing direction is not exactly orthogonal to the
driving direction, some of the energy gets coupled to the sensing
direction''~\cite{khazaaleh_vulnerability_2019}. This causes a false reading,
which is typically corrected ``by employing a demodulator in
the readout circuit''~\cite{khazaaleh_vulnerability_2019}.
When the transmission frequency is slightly different from the sensing frequency,
the gyroscope generates ``measurable output'', whose frequency equals ``the
difference between the driving frequency and the frequency of the acoustic
signal''~\cite{khazaaleh_vulnerability_2019}. As shown experimentally,
this model also explains why it is better to transmit near the resonant
frequency rather than exactly at it~\cite{khazaaleh_vulnerability_2019}.
It also suggests that low-pass filters or differential measurements through additional
proof masses are ineffective countermeasures against out-of-band acoustic signal
injection attacks~\cite{khazaaleh_vulnerability_2019}.

Although Tu et al.\ succeeded in controlling gyroscopes in phones, scooters,
stabilizers, screwdrivers, and VR headsets among others, only DoS attacks
were successful against accelerometers~\cite{tu_injected_actuation_2018}.
MEMS accelerometers consist of spring-mass systems, so acceleration results
in a deflection of the seismic mass. This deflection ``is detected by means of
capacitive elements, the capacitances
of which change with deflection, or by piezoresistive elements that detect
strain induced by the motion of the seismic mass through a change in
resistor values''~\cite{laermer_mechanical_2006}. Acoustic vibrations at the
resonant frequencies of the spring-mass systems can also displace the suspended
mass, making them vulnerable to out-of-band attacks. According to Trippel et al.,
insecure amplifiers and low-pass filters (LPFs) prior to the accelerometer
ADCs can demodulate both Amplitude-Modulated (AM)
and Phase-Modulated ($\phi$M) attacker injections~\cite{trippel_walnut_2017}. These
insecurities are the results of clipping non-linearities and permissive filtering
respectively, and allow for both biasing and control attacks.

Trippel et al.\ used accelerometers to spell words,
naming their work ``WALNUT'' for the output of the spoofed
sensor measurements. They were also able to control off-the-shelf devices,
such as remote-controlled (RC) cars, and Fitbit fitness tracking
wristbands~\cite{trippel_walnut_2017}. Although spoofing step counts might
seem innocuous, companies often offer financial rewards for health-related
activity~\cite{trippel_walnut_2017}, so cheating devices (which
do not yet exploit out-of-band effects) are already being sold~\cite{xue_gadget_2018}.
Most attacks by Trippel et al.~\cite{trippel_walnut_2017} were performed at
distances of $\SI{10}{\centi\meter}$, with a speaker producing an
SPL of $\SI{110}{\deci\bel}$. The duration of control over the output
of the MEMS sensors was often limited to $\SIrange{1}{2}{\second}$
(and up to $\SI{30}{\second}$) due to sampling rate drifts.
Moreover, it was shown that the three axes do not behave identically to
acoustic injections. For example,
there are some MEMS devices for which only the $x$-axis responds to acoustic transmissions,
while others are vulnerable in all three axes, but at different resonant frequencies.

Although Trippel et al.~\cite{trippel_walnut_2017} attacked a single sensor
in one direction at a time, Nashimoto et al.'s ``Sensor CON-Fusion''
investigated whether {\em sensor fusion} using a Kalman Filter can
improve the robustness of measurements~\cite{nashimoto_sensor_2018}.
It was shown that ``while sensor fusion introduces a certain degree of
attack resilience, it remains susceptible'' to combined
acoustic and electromagnetic injections~\cite{nashimoto_sensor_2018}.
Specifically, Nashimoto succeeded in simultaneously controlling the roll,
pitch, and yaw (the three angular axes in aircraft nomenclature) by fusing the
outputs of an accelerometer, a gyroscope, and magnetometer. However,
in non-simulated environments, ``there is an error in the roll angle'', and
``the resulting inclination does not last long''~\cite{nashimoto_sensor_2018}.
Although fusion is further explored in the context of defense mechanisms
(Section~\ref{sec:defenses}), the above discussion leads to the following
research question:

\begin{open}
Is it possible to use acoustic injections to precisely control MEMS gyroscope and
accelerometer measurements in all three directions simultaneously and/or for
longer periods of time?
\end{open}

In a different strand of research, ten years after a video
demonstrating that shouting in a data center
causes unusually high disk I/O latency \cite{gregg_unusual_2008}, Shahrad
et al.\ showed that acoustic transmissions can cause vibrations
in Hard-Disk Drives (HDDs)~\cite{shahrad_acoustic_2018}.
These vibrations result in read and write errors at distances up to $\SI{70}{\centi\meter}$
with a sound level of $\SI{102.6}{\deci\bel A}$~\cite{shahrad_acoustic_2018}.
As a result, they can make systems unresponsive, even leading to
Blue Screen of Death (BSOD) errors~\cite{shahrad_acoustic_2018}.

Although Shahrad et al.\ primarily focused on the effect of
the angle of transmission~\cite{shahrad_acoustic_2018}, research
conducted in parallel more precisely pinpointed the root cause of the issue
using Finite Element Analysis~\cite{bolton_blue_2018}. Specifically, it was
shown that (audible) acoustic waves ``can displace a read/write head or disk
platter outside of operational bounds, inducing throughput loss'', even
though the displacement is of only a few nanometers~\cite{bolton_blue_2018}.
In addition, in their ``Blue Note'' paper, Bolton et al.\ also
used ultrasonic transmissions to attack the shock sensors which are
meant to protect HDDs during sudden drops by ``parking
the read/write head''~\cite{bolton_blue_2018}:
modern HDDs contain ``piezo shock sensors or MEMS capacitive accelerometers''
to ``detect sudden disturbances''~\cite{bolton_blue_2018}, and can also
be attacked through ultrasound transmissions at their resonant frequencies.
Through these malicious acoustic attacks, SPL levels of up to $\SI{130}{\deci\bel}$
cause HDDs at distances of $\SI{10}{\centi\meter}$ to become unresponsive,
thus disabling laptops and video recorders~\cite{bolton_blue_2018}.

\begin{figure}[t!]
    \centering
    \includegraphics[width=\linewidth]{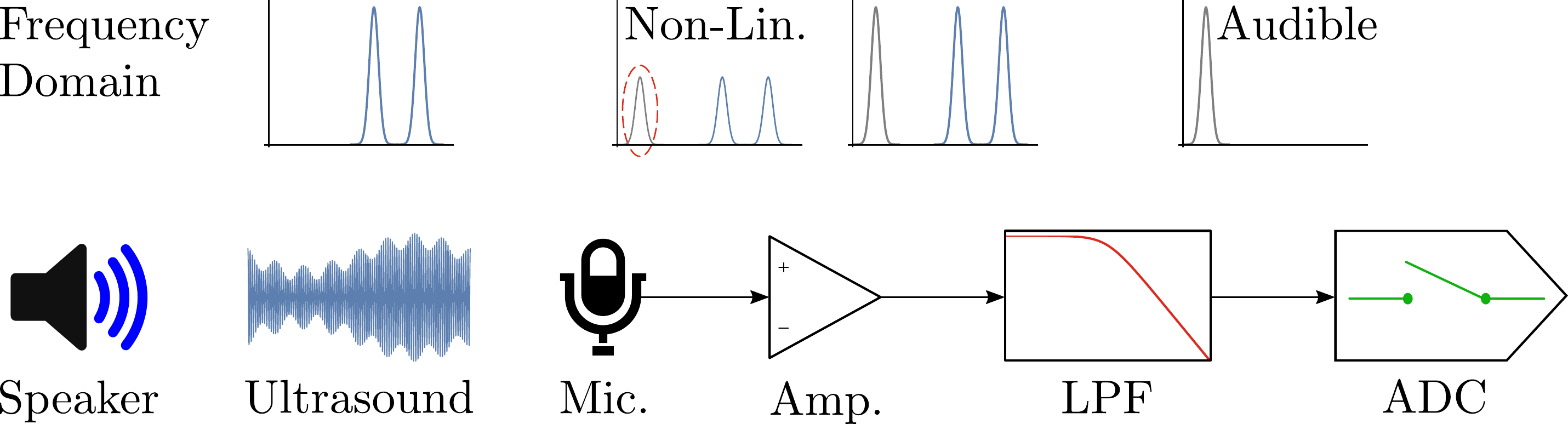}
    \caption{High-level overview of ultrasonic attacks against
             microphones~\cite{song_inaudible_2017, zhang_dolphin_2017,
             roy_inaudible_2018, yan_feasibility_2019}.
             A speaker transmits inaudible tones, but intermodulation products
             are produced due to non-linearities in the microphone
             and amplifier. A Low-Pass Filter (LPF) removes ultrasound frequencies, so
             the Analog-to-Digital Converter (ADC) records only audible by-products.}
     \label{fig:acoustic}
\end{figure}

Not all acoustic attacks transmit at resonant frequencies. By contrast, other
research targets microphone non-linearities to cause inaudible sound to be
recorded. Microphones operate by converting the mechanical deformation of a
membrane (caused by the air pressure of a sound wave) into a
capacitive change, which produces an Alternating Current (AC)
signal~\cite{zhang_dolphin_2017}. This process also has
non-linearities which produce second-order components~\cite{zhang_dolphin_2017},
including harmonics and intermodulation products, as discussed in Section~\ref{sec:model}.

Early work on acoustic attacks primarily focused on adversarial control of
machine learning in speech recognition systems~\cite{vaidya_cocaine_2015,
carlini_hidden_2016}. Such research did not take advantage of non-linearities,
and was in-band, as the transmissions were audible (although indecipherable
by humans). However, later investigations revolved around ensuring that the
transmitted frequencies are beyond the human-audible
range ($\SI{20}{\kilo\hertz}$). This was first accomplished by Zhang
et al.\ in their ``DolphinAttack'' paper, which exploited non-linearities in
microphone sensors~\cite{zhang_dolphin_2017}, as shown in Figure~\ref{fig:acoustic}.

The work by Zhang et al.\ showed how to transform (modulated) ultrasound
transmissions into valid commands which were executed by speech recognition
systems such as Apple Siri, Google Now, Microsoft Cortana, and
Amazon Alexa~\cite{zhang_dolphin_2017}. The same authors later
expanded on their attack by testing different setups, and increasing the
attack distance from $\SI{1.7}{\meter}$ to $\SI{19.8}{\meter}$~\cite{yan_feasibility_2019}.
This was done by replacing the $\SI{125}{\deci\bel}$ source with an ultrasonic
transmitter array and amplifier outputting $\SI{1.5}{\watt}$ to increase sound
pressure. Above this transmission power, the attack becomes audible due to
non-linearities in the transmission medium (air) and the source speakers~\cite{zhang_dolphin_2017}.

Earlier, Roy et al.\ had also noted that non-linearities in speakers make it
harder for an adversary to increase the attack distance:
 ``increasing the transmit power at the speaker triggers non-linearities
at the speaker's own diaphragm and amplifier, resulting in an audible
[output]''~\cite{roy_inaudible_2018}. Instead, multiple
speakers in the form of an ultrasonic speaker array can be used to
attack voice recognition systems including Amazon Alexa and Google Now
at a distance of up to $\SI{7.6}{\meter}$
using a $\SI{6}{\watt}$ source~\cite{roy_inaudible_2018}.
The attack works by partitioning the audio spectrum across the various speakers
in a way that ``reduces the audible leakage from any given speaker''
while minimizing the total leakage power~\cite{roy_inaudible_2018}.
This prevents any of the non-linearities (and the transmitted
signal itself) from being audible. It should be noted that if multiple
non-cooperating ultrasonic sources are emitting simultaneously, intermodulation
distortions can create audible byproducts~\cite{yan_cuba_2019}, allowing for
the detection of potential attacks.

Parallel to the 2017 DolphinAttack paper~\cite{zhang_dolphin_2017},
similar research was in progress at Princeton~\cite{song_inaudible_2017}:
Song and Mittal also succeeded in injecting inaudible voice commands to an
Amazon Echo and an Android phone. Although they accomplished relatively long
distances ($\SI{3.54}{\meter}$ with an input power of $\SI{23.7}{\watt}$),
their work remains in poster format. Moreover,
prior to their inaudible voice commands work~\cite{roy_inaudible_2018},
Roy et al.\ also used inaudible ultrasound transmissions
to record audible sounds. Specifically, in their ``BackDoor'' work, they proposed
a high-bandwidth covert channel that operates at up to $\SI{1.5}{\meter}$
using a $\SI{2}{\watt}$ source~\cite{roy_backdoor_2017}.
Although covert channels are not discussed in this survey, the work by Roy
et al.\ is included due to the methodology which naturally led to their later work.

More concretely, instead of using amplitude modulation over a single
frequency like Zhang et al.~\cite{zhang_dolphin_2017}, Roy et al.\ simultaneously played
two ultrasound tones whose shadows create audible sounds (only sensed by
microphones) due to non-linearities~\cite{roy_backdoor_2017}. They showed that amplitude
modulation could not be used due to non-linearities in the ultrasound transmitters
themselves, which would result in audible signals. Instead, further pre-computation
was required to remove the ``ringing effect'', where ``the transmitted sound becomes slightly
audible even with FM modulation''~\cite{roy_backdoor_2017}. These results
point towards the next open question~\cite{yan_feasibility_2019}:

\begin{open}
How can non-linear acoustics in the transmission
medium (air and speakers) be avoided to further extend the range of inaudible attacks?
\end{open}

\section{Optical and Thermal Manipulations}
\label{sec:others}

Although electromagnetic, conducted, and acoustic attacks form the majority
of out-of-band signal injection attacks, there has been some research on
optical attacks, as well as temperature attacks which bias RO-based TRNGs.
In the former category, attacks exploit permissive filtering and poor shielding in
interfaces which only expect ambient environmental conditions. Most papers so far
have targeted sensors in an in-band fashion: out-of-scope
research includes attacks on sonars, radars, and
LiDARs~\cite{petit_remote_2015, yan_autonomous_vehicles_2016, shin_illusion_2017,
xu_analyzing_2018}, as well as visible-light attacks on cameras in unmanned aerial vehicles
(UAVs)~\cite{davidson_controlling_2016} and cars~\cite{petit_remote_2015,
yan_autonomous_vehicles_2016, xu_analyzing_2018}.

Researchers have also hypothesized that excessive light injections would
blind car cameras and confuse auto-controls in automated
vehicles~\cite{petit_potential_2015}. Indeed, limited success against cameras has been
achieved using Ultraviolet (UV) and Infrared (IR) lasers
up to $\SI{2}{\meter}$ away~\cite{petit_remote_2015}.
However, attacks were only possible in dark environments, and the results
were not reproducible with invisible lasers against other makes and models of
CMOS cameras~\cite{yan_autonomous_vehicles_2016}.

In a different strand of research, Park et al.\ showed that some
medical infusion pumps are not well-protected
against adversarial optical injections~\cite{park_pump_2016}. Specifically,
in order to measure the flow rate of the medicine being administered, pumps are
fitted with drip measurement sensors. These sensors consist of an IR emitter and receiver
facing each other. When a drop passes through the sensor, the IR receiver
temporarily senses less light due to diffusion, allowing for the rate to be measured.
However, because the sensor is not well-enclosed, an adversary can shine
an IR laser into the sensor, causing these drops to be undetected,
thereby saturating the sensor. By then un-blinding the sensor, the attackers
can also trick the firmware into detecting fake drops, and bypass alarms.

Adversaries can therefore selectively both over- and under-infuse
a patient for an extended period of time, and for a variety of normal flow volumes.
However, this can only be done with coarse-grained control over the real flow rate.
Most of the experiments by Park et al.\ were conducted at a distance of
$\SI{10}{\centi\meter}$. Success was nevertheless reported up to $\SI{12}{\meter}$ away
using a $\SI{30}{\milli\watt}$ IR laser pointer~\cite{park_pump_2016}.
These results show that optical attacks can reach meaningful distances, but
are, of course, limited by line-of-sight considerations.

It should be noted that the attack by Park et al.\ should be considered to be
out-of-band, as the pump was not meant to receive external stimuli, in contrast
to, for example, LiDARs, which are supposed to interact with external objects.
In other words, a LiDAR depends on its surroundings to reflect its
transmitted pulses and therefore infer the distance to the interfering objects.
On the other hand, the drip sensor and (part of) the intravenous (IV) tube
could be enclosed and shielded from
the environment. This naturally leads to the defense mechanisms proposed by
Park et al.~\cite{park_pump_2016}, which are discussed in Section~\ref{sec:defenses},
and which beg the following question:

\begin{open}
Are all out-of-band optical attacks a matter of poor filtering and shielding?
\end{open}

\begin{figure}[t!]
    \centering
    \includegraphics[width=.7\linewidth]{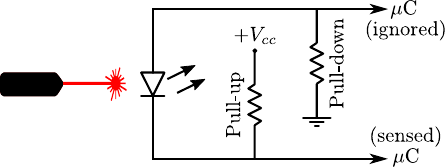}
    \caption{Unconventional circuit layout to convert a Light Emitting Diode (LED)
             into a photodiode~\cite{loughry_oops_2019}. Pull-up and -down
             resistors can be internal to microcontroller input pins.}
     \label{fig:optical}
\end{figure}

Under somewhat unrealistic assumptions, the answer to the above question might
be ``no''. In their typical mode of operation, Light Emitting Diodes (LEDs)
convert current to light. However, they can also function in the reverse
by producing current when illuminated~\cite{mims_bidirectional_1979, mims_sun_1992}.
In preliminary experiments, Loughry recently showed that this behavior
can be exploited for an optical covert channel~\cite{loughry_oops_2019}.
The setup is certainly unconventional: both LED pins are connected to
microcontroller GPIO pins, which are configured as inputs with (internal) pull-up
and pull-down resistors, as shown in Figure~\ref{fig:optical}.
According to Loughry, seven out of ten LEDs tested responded to laser and light of
different wavelengths, with some LEDs producing measurable current at
both ends (anode and cathode)~\cite{loughry_oops_2019}. Whether out-of-band
signal injections or fault attacks can exploit this effect is yet
to be seen~\cite{loughry_oops_2019}.

The final class of attacks exploits the dependence of ring
oscillators (ROs) on temperature to reduce the entropy of the TRNG. It is only mentioned
here for completeness, as distance requirements would dictate physical
access to the device under attack. Early work in the area showed that
statistical randomness tests of RO-based TRNGs would fail for certain FPGA
temperatures~\cite{simka_active_2006}. More detailed experimentation
conducted a few years later using different heat-transfer
methods (resistor heater, Peltier cooler, and liquid nitrogen) then showed
that ``the hotter the temperature, the larger the bias''~\cite{soucarros_influence_2011}.
Martín et al.~\cite{martin_fault_2015} also investigated the effect of temperature across
multiple TNRG designs based on Self-Timed Rings (STRs), which do not
exhibit the frequency locking effects discussed earlier~\cite{cherkaoui_self_2013}.
It was shown that the effects of temperature increases on
the randomness of STR-based TRNGs were not significant, due to a combination of
a decrease in frequency and an increase in jitter due to thermal noise~\cite{martin_fault_2015}.

Since devices were operating within their specifications, these biasing effects could
be considered out-of-band attacks which operate at limited distances. This is in contrast
to, for example, Martín et al.'s work which investigates the entropy
of TRNGs in response to ionizing radiation~\cite{martin_entropy_2018}. However,
remote conducted attacks on TRNGs are possible using local voltage
drops~\cite{mahmoud_timing_2019}. Whether it is possible to reproduce these
effects using heating circuits~\cite{agne_seven_2014} remains an open question:

\begin{open}
Can software-only thermal effects bias RO-based TRNGs in multi-tenant FPGAs?
\end{open}

Overall, the efficacy of out-of-band optical thermal injections
has been limited so far, due to the limited nature of vulnerable interfaces
and proximity considerations respectively.

\begin{landscape}
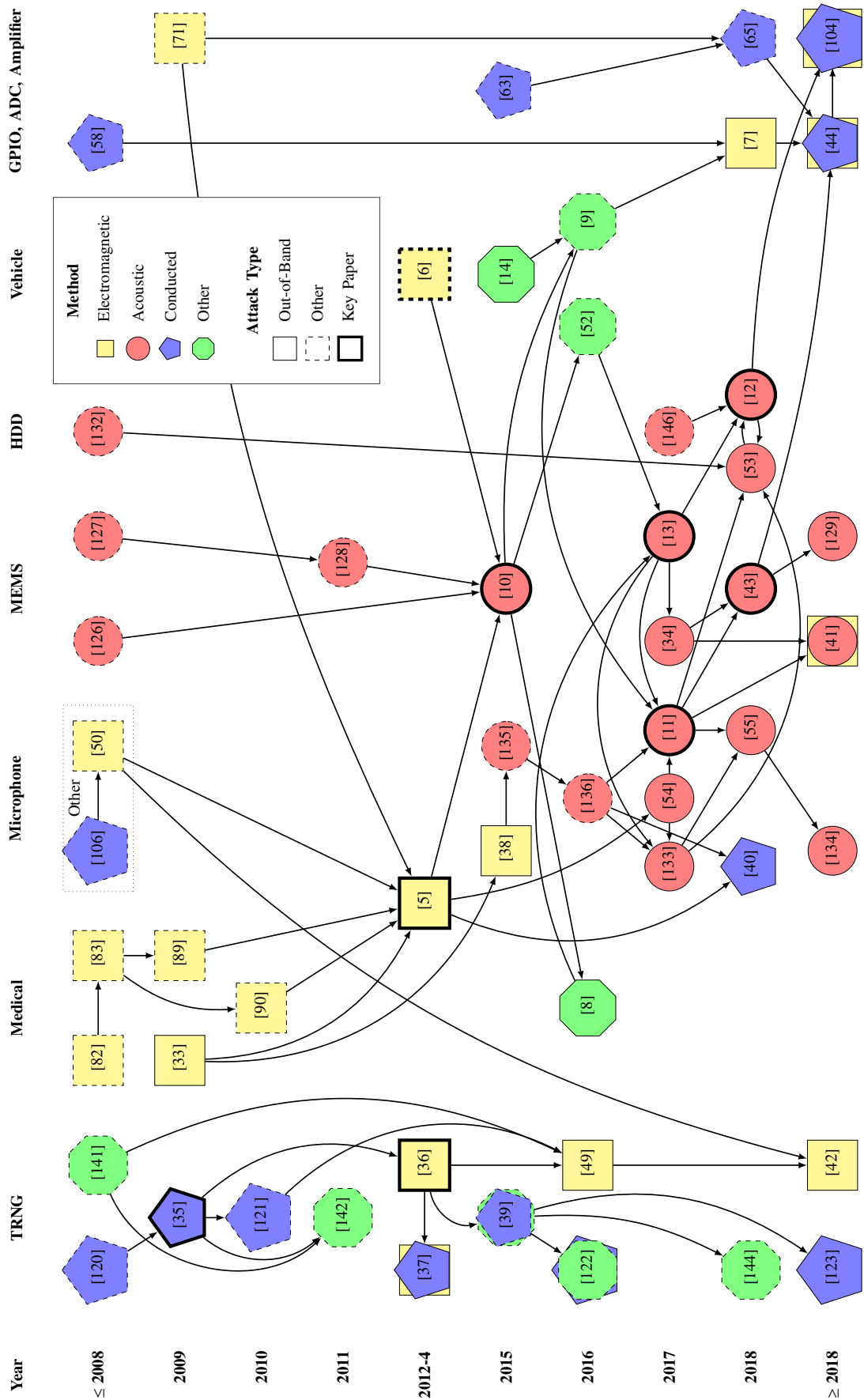
\begin{figure}
\centering
\resizebox{1.35\textwidth}{!}{
\tikzstyle{highlight} = [line width=2pt]
\tikzstyle{label} = [rectangle,draw=none,fill=none, minimum size=1.62cm, align=center, inner sep=0pt]
\tikzstyle{acoustic} = [circle, minimum size=1.05cm, align=center, draw=black, fill=red!50, inner sep=0pt]
\tikzstyle{em} = [regular polygon,regular polygon sides=4, minimum size=1.52cm, align=center, draw=black, fill=yellow!50, inner sep=0pt]
\tikzstyle{conducted} = [regular polygon,regular polygon sides=5, minimum size=1.3cm, align=center, draw=black, fill=blue!50, inner sep=0pt]
\tikzstyle{others} = [regular polygon,regular polygon sides=8, minimum size=1.3cm, align=center, draw=black, fill=green!50, inner sep=0pt]
\begin{tikzpicture}[x=2.25cm, y=1.75cm, on grid]
        \node (year) [label] {\textbf{Year}};
        \node (2006) [label, below = 1 of year] {$\mathbf{\le2008}$};
        \node (2009) [label, below = 1 of 2006] {\quad\textbf{2009}};
        \node (2010) [label, below = 1 of 2009] {\quad\textbf{2010}};
        \node (2011) [label, below = 1 of 2010] {\quad\textbf{2011}};
        \node (2013) [label, below = 1 of 2011] {\textbf{2012-4}};
        \node (2015) [label, below = 1 of 2013] {\quad\textbf{2015}};
        \node (2016) [label, below = 1 of 2015] {\quad\textbf{2016}};
        \node (2017) [label, below = 1 of 2016] {\quad\textbf{2017}};
        \node (2018) [label, below = 1 of 2017] {\quad\textbf{2018}};
        \node (2019) [label, below = 1 of 2018] {$\mathbf{\ge2018}$};

        \node (trng) [label, right = 1.5 of year] {\textbf{TRNG}};
        \node (mesgarzadeh) [conducted, right = 1 of 2006, dashed] {\cite{mesgarzadeh_study_2005}};
        \node (simka) [others, right = 2 of 2006, dashed] {\cite{simka_active_2006}};
        \node (markettos) [conducted, highlight, right = 1.5 of 2009] {\cite{markettos_trng_injection_2009}};
        \node (bochard) [conducted, right = 1.5 of 2010, dashed] {\cite{bochard_true_2010}};
        \node (soucarros) [others, right = 1.5 of 2011, dashed] {\cite{soucarros_influence_2011}};
        \node (bayon2012) [em, highlight, right = 2 of 2013] {\cite{bayon_trng_em_2012}};
        \node (buchovecka) [em, right = 1 of 2013] {\cite{buchovecka_frequency_2013}};
        \node (buchovecka_cond) [conducted, right = 1 of 2013] {\cite{buchovecka_frequency_2013}};
        \node (martin2015) [others, dashed, right = 1.5 of 2015] {\cite{martin_fault_2015}}; 
        \node (martin2015_cond) [conducted, dashed, right = 1.5 of 2015] {\cite{martin_fault_2015}};
        \node (cao_cond) [conducted, right = 1 of 2016] {\cite{cao_exploring_2016}};
        \node (cao) [others, right = 1 of 2016, dashed] {\cite{cao_exploring_2016}}; 
        \node (bayon2016) [em, right = 2 of 2016] {\cite{bayon_fault_2016}};
        \node (martin2018) [others, right = 1 of 2018, dashed] {\cite{martin_entropy_2018}};
        \node (osuka) [em, right = 2 of 2019] {\cite{osuka_em_2019}};
        \node (mahmoud) [conducted, right = 1 of 2019] {\cite{mahmoud_timing_2019}};

        \node (medical) [label, right = 2 of trng] {\textbf{Medical}};
        \node (irnich) [em, dashed, right = 3 of 2006] {\cite{irnich_interference_1996}};
        \node (hayes) [em, dashed, right = 4 of 2006] {\cite{hayes_interference_1997}};
        \node (rasmussen) [em, right = 3 of 2009] {\cite{rasmussen_proximity_based_2009}};
        \node (lee) [em, dashed, right = 4 of 2009] {\cite{lee_clinically_2009}};
        \node (seidman) [em, dashed, right = 3.5 of 2010] {\cite{seidman_vitro_2010}};
        \node (kune) [em, highlight, right = 4.5 of 2013] {\cite{kune_ghost_2013}};
        \node (park) [others, right = 3.5 of 2016] {\cite{park_pump_2016}};

        \node (mic) [label, right = 2 of medical] {\textbf{Microphone}};
        \node (mic_rect) [rectangle, dotted, draw=black!75, minimum width=4cm, minimum height=1.5cm, below = 1 of mic, inner sep=0em]{};
        \node [anchor=north] at (mic_rect.north) {Other};
        \node (parfenov) [conducted, dashed, right = 5 of 2006] {\cite{parfenov_conducted_2004}};
        \node (radasky) [em, dashed, right = 6 of 2006] {\cite{radasky_introduction_2004}};
        \node (kasmi) [em, right = 5 of 2015] {\cite{kasmi_iemi_2015}};
        \node (vaidya) [acoustic, right = 6 of 2015, dashed] {\cite{vaidya_cocaine_2015}};
        \node (carlini) [acoustic, right = 5.5 of 2016, dashed] {\cite{carlini_hidden_2016}};
        \node (song) [acoustic, right = 4.85 of 2017] {\cite{song_inaudible_2017}};
        \node (roy2017) [acoustic, right = 5.5 of 2017] {\cite{roy_backdoor_2017}};
        \node (zhang) [acoustic, highlight, right = 6.15 of 2017] {\cite{zhang_dolphin_2017}};
        \node (esteves) [conducted, right = 4.85 of 2018] {\cite{esteves_remote_2018}};
        \node (roy2018) [acoustic, right = 6.15 of 2018] {\cite{roy_inaudible_2018}};
        \node (yan2019) [acoustic, right = 5 of 2019] {\cite{yan_feasibility_2019}};

        \node (mems) [label, right = 2 of mic] {\textbf{MEMS}};
        \node (castro) [acoustic, dashed, right = 7 of 2006] {\cite{castro_influence_2007}};
        \node (dean2007) [acoustic, dashed, right = 8 of 2006] {\cite{dean_degradation_2007}};
        \node (dean2011) [acoustic, dashed, right = 7.75 of 2011] {\cite{dean_characterization_2011}};
        \node (son) [acoustic, highlight, right = 7.5 of 2015] {\cite{son_rocking_2015}};
        \node (wang) [acoustic, right = 7 of 2017] {\cite{wang_sonic_2017}};
        \node (trippel) [acoustic, highlight, right = 8 of 2017] {\cite{trippel_walnut_2017}};
        \node (tu) [acoustic, highlight, right = 7.5 of 2018] {\cite{tu_injected_actuation_2018}};
        \node (nashimoto_em) [em, right = 7 of 2019] {\cite{nashimoto_sensor_2018}};
        \node (nashimoto_acoustic) [acoustic, right = 7 of 2019] {\cite{nashimoto_sensor_2018}};
        \node (khazaaleh) [acoustic, right = 8 of 2019] {\cite{khazaaleh_vulnerability_2019}};

        \node (hdds) [label, right = 1.5 of mems] {\textbf{HDD}};
        \node (gregg) [acoustic, right = 9 of 2006, dashed] {\cite{gregg_unusual_2008}};
        \node (dutta) [acoustic, right = 9 of 2017, dashed] {\cite{dutta_performance_2017}};
        \node (bolton) [acoustic, highlight, right = 9.35 of 2018] {\cite{bolton_blue_2018}};
        \node (shahrad) [acoustic, right = 8.65 of 2018] {\cite{shahrad_acoustic_2018}};

        \node (vehicles) [label, right = 1.5 of hdds] {\textbf{Vehicle}};
        \node (shoukry) [em, highlight, right = 10.5 of 2013, dashed] {\cite{shoukry_non_invasive_2013}};
        \node (petit) [others, right = 10.5 of 2015] {\cite{petit_remote_2015}};
        \node (yan) [others, right = 11 of 2016, dashed] {\cite{yan_autonomous_vehicles_2016}};
        \node (davidson) [others, right = 10 of 2016, dashed] {\cite{davidson_controlling_2016}};

        \node (gpio) [label, right = 1.75 of vehicles] {\textbf{GPIO, ADC, Amplifier}};
        \node (boyer) [conducted, right = 11.75 of 2006, dashed] {\cite{boyer_modeling_power_injection_2007}};
        \node (redoute) [em, right = 12.75 of 2009, dashed] {\cite{redoute_emc_2009}};
        \node (ayed) [conducted, dashed, right = 12.25 of 2015] {\cite{ayed_immunity_modeling_adc_2015}};
        \node (selvaraj) [em, right = 11.75 of 2018] {\cite{selvaraj_induction_embedded_2018}};
        \node (kennedy) [conducted, dashed, right = 12.75 of 2018] {\cite{kennedy_flash_adc_emi_2018}};
        \node (giechaskiel) [em, right = 11.75 of 2019]{\cite{giechaskiel_framework_2019}};
        \node (giechaskiel_cond) [conducted, right = 11.75 of 2019]{\cite{giechaskiel_framework_2019}};
        \node (tu2019) [em, right = 12.75 of 2019]{\cite{tu_trick_2019}};
        \node (tu2019_cond) [conducted, right = 12.75 of 2019]{\cite{tu_trick_2019}};

        \gpath{mesgarzadeh}{markettos}
        \gpath[-55]{simka}{soucarros}
        \gpath[25]{simka}{bayon2016}
        \gpath{markettos}{bochard}
        \gpath[-40]{markettos}{soucarros}
        \gpath[30]{markettos}{bayon2012}
        \gpath[35]{bochard}{bayon2016}
        \gpath{bayon2012}{buchovecka_cond}
        \gpath[-45]{bayon2012}{martin2015}
        \gpath{bayon2012}{bayon2016}
        \gpath{martin2015}{cao}
        \gpath[15]{martin2015}{martin2018}
        \gpath[25]{martin2015}{mahmoud}
        \gpath{bayon2016}{osuka}

        \gpath{irnich}{hayes}
        \gpath{hayes}{lee}
        \gpath[-20]{hayes}{seidman}
        \gpath[-30]{rasmussen}{kune}
        \gpath[-35]{rasmussen}{kasmi}
        \gpath{kasmi}{vaidya}
        \gpath{lee}{kune}
        \gpath{seidman}{kune}
        \gpath[-15]{kune}{roy2017}
        \gpath{kune}{son}
        \gpath[-30]{kune}{esteves}
        \gpath[35]{park}{trippel}

        \gpath{parfenov}{radasky}
        \gpath{radasky}{kune}
        \gpath[-15]{radasky}{osuka}
        \gpath{vaidya}{carlini}
        \gpath{carlini}{zhang}
        \gpath{carlini}{song}
        \gpath{carlini}{esteves}
        \gpath{roy2017}{zhang}
        \gpath{roy2017}{song}
        \gpath[-40]{song}{shahrad}
        \gpath{song}{roy2018}
        \gpath{zhang}{roy2018}
        \gpath{zhang}{shahrad}
        \gpath{zhang}{tu}
        \gpath{roy2018}{yan2019}

        \gpath{castro}{son}
        \gpath{dean2007}{dean2011}
        \gpath{dean2011}{son}
        \gpath{son}{park}
        \gpath[15]{son}{yan}
        \gpath{son}{davidson}
        \gpath{trippel}{wang}
        \gpath[-25]{trippel}{zhang}
        \gpath[-40]{trippel}{song}
        \gpath{trippel}{bolton}
        \gpath{zhang}{nashimoto_em}
        \gpath{wang}{nashimoto_em}
        \gpath{wang}{tu}
        \gpath[-6]{tu}{giechaskiel}
        \gpath{tu}{khazaaleh}

        \gpath{gregg}{shahrad}
        \gpath{dutta}{bolton}
        \gpath[15]{bolton}{shahrad}
        \gpath[15]{shahrad}{bolton}
        \gpath[10]{bolton}{tu2019}

        \gpath{shoukry}{son}
        \gpath{petit}{yan}
        \gpath{yan}{selvaraj}
        \gpath[-30]{yan}{zhang}
        \gpath{davidson}{trippel}

        \gpath{boyer}{selvaraj}
        \gpath[-10]{redoute}{kune}
        \gpath{redoute}{kennedy}
        \gpath{ayed}{kennedy}
        \gpath{kennedy}{giechaskiel}
        \gpath{selvaraj}{giechaskiel_cond}
        \gpath{giechaskiel}{tu2019}

        \node (legend_rect) [rectangle, draw=black!75, fill=white, xshift=-1em, yshift=-1em, minimum width=4cm, minimum height=7cm, below = 2.25 of vehicles]{};
        \node (legend_attack) [below = 0.1 of legend_rect.north, anchor=north] {\textbf{Method}};

        \node (em) [em, below = 0.4 of legend_attack.west, xshift=-1em, anchor=east, minimum size=0.5cm, align=left] {};
        \node (em_label)[right = 0.2 of em, align = left, anchor=west] {Electromagnetic};

        \node (acoustic) [acoustic, below = 0.4 of em, minimum size=0.5cm] {};
        \node (acoustic_label)[right = 0.2 of acoustic, anchor=west, align = left] {Acoustic};

        \node (conducted) [conducted, below = 0.4 of acoustic, minimum size=0.5cm] {};
        \node (conducted_label) [right = 0.2 of conducted, anchor=west, align = left] {Conducted};

        \node (others) [others, below = 0.4 of conducted, minimum size=0.5cm] {};
        \node (others_label) [right = 0.2 of others, anchor=west, align = left] {Other};

        \node (legend_box) [below = 2.3 of legend_rect.north, anchor=north]{\textbf{Attack Type}};
        \node (regular) [draw=black, below = 1 of others, minimum size=0.5cm] {};
        \node (regular_label) [right = 0.2 of regular, anchor=west, align = left] {Out-of-Band};

        \node (other) [draw=black, dashed, below = 0.4 of regular, minimum size=0.5cm] {};
        \node (other_label) [right = 0.2 of other, anchor=west, align = left] {Other};

        \node (important) [draw=black, ultra thick, below = 0.4 of other, minimum size=0.5cm] {};
        \node (important_label) [right = 0.2 of important, anchor=west, align = left] {Key Paper};

\end{tikzpicture}
}
\caption{Evolutionary and thematic taxonomy of out-of-band signal injection attacks, categorized by
        topic and methodology. Cross-influences between out-of-band signal injections
        and some key related works are also shown. Highly influential papers are discussed
        more extensively in Section~\ref{sec:taxonomy}.}
\label{fig:attacks}
\end{figure}
\end{landscape}

\section{Taxonomy of Attacks}
\label{sec:taxonomy}

This section presents a taxonomy of out-of-band signal injection attacks,
tracing their evolution through time and topic, and identifying commonalities
in their methodology and source of vulnerability. We first
highlight thematic and evolutionary cross-influences between
the various works studied in this survey (Figure~\ref{fig:attacks}),
and then categorize the key hardware imperfections
that make them possible (Table~\ref{table:attacks}).
We start with a citation graph of out-of-band attacks and related work in
Figure~\ref{fig:attacks}, where an edge $X\rightarrow Y$ indicates that
$X$ is cited by $Y$. To reduce clutter, if a paper $X$ is cited
by both $Y$ and $Z$, and $Y$ is cited by $Z$, then no
arrow from $X$ to $Y$ is drawn.

\begin{table*}[!t]
    \centering
    \caption{Sources of vulnerability, methods, and effects for
             out-of-band signal injection attacks along with maximum power and distance.
             For each potential source of vulnerability, \yes{} signifies that the
             paper claims the vulnerability contributes to the attack,
             while \no{} that it does not.
             The attack methods used are
             \acoustic{}~acoustic, \conducted{}~conducted,
             \electromagnetic{}~electromagnetic, and \optical{}~optical.
             The effects achieved include \protect\disruption{}\protect~disruption, \protect\bias{}\protect~bias, and
             \protect\control{}\protect~control of the output, while \protect\theoretical{}\protect~denotes a
             theoretical or partially-realized attack.
             Resonance (\textdagger{}) includes frequency locking for ring
             oscillators, and unintentional wire antennas for EM attacks.}
    \begin{tabular}{@{}rrrrrrccccccc@{}}
    \\\toprule
        \textbf{Authors} & \textbf{Year} & \textbf{Ref.} & \textbf{Target} &
        \textbf{Power} & \textbf{Distance} &\rot{\textbf{Method}} &
        \rot{\textbf{Effect}} &\rot{\textbf{Resonance (\textdagger{})}} &
        \rot{\textbf{Non-Linearity}} & \rot{\textbf{Improper Filtering}} &
        \rot{\textbf{Poor Shielding}} & \rot{\textbf{Insecure Algorithm}}\\
    \midrule
    Rasmussen et al. & 2009 & \cite{rasmussen_proximity_based_2009} & Medical & (unspecified) & (unspecified) & \electromagnetic{} & \theoretical{} & \yes & \no & \no & \yes & \no \\
    Petit et al. & 2015 & \cite{petit_remote_2015} & Camera & (unspecified) & $\SI{2.00}{\meter}$ & \optical{} & \disruption{} & \no & \no & \yes & \no & \no \\
    Cao et al. & 2016 & \cite{cao_exploring_2016} & TRNG & (under-volt) & (direct) & \conducted{} & \bias{} & \yes & \no & \no & \no & \no \\
    Wang et al. & 2017 & \cite{wang_sonic_2017} & MEMS & (unspecified) & (unspecified) & \acoustic{} & \control{} & \yes & \yes & \yes & \yes & \no \\
    Nashimoto et al. & 2018 & \cite{nashimoto_sensor_2018} & MEMS & (unspecified) & (unspecified) & \acoustic{}\electromagnetic{} & \control{} & \yes & \no & \yes & \yes & \yes \\
    Mahmoud and Stoilovi\'{c} & 2019 & \cite{mahmoud_timing_2019} & TRNG & (140K ROs) & (internal) & \conducted{} & \bias{} & \yes & \no & \no & \no & \no \\
    \midrule
    Markettos and Moore & 2009 & \cite{markettos_trng_injection_2009} & TRNG & $\SI{0.002}{\watt}$ & (direct) & \conducted{} & \bias{} & \yes & \no & \yes & \no & \yes \\
    Bayon et al. & 2012 & \cite{bayon_trng_em_2012} & TRNG & $\SI{0.003}{\watt}$ & $\SI{100}{\textcolor{black}{\textbf{\micro}}\meter}$ & \electromagnetic{} & \bias{} & \yes & \no & \no & \no & \no \\
    Buchovecka et al. & 2013 & \cite{buchovecka_frequency_2013} & TRNG & $\le\SI{0.563}{\watt}$ & ``near'' & \conducted{}\electromagnetic{} & \bias{} & \yes & \no & \no & \no & \no \\
    Kune et al. & 2013 & \cite{kune_ghost_2013} & Med. \& Mic. & $\SI{10.000}{\watt}$ & $\SI{1.67}{\meter}$ & \electromagnetic{} & \control{} & \yes & \yes & \yes & \yes & \no \\
    Kasmi and Esteves & 2015 & \cite{kasmi_iemi_2015} & Microphone & $\SI{200.000}{\watt}$ & $\SI{4.00}{\meter}$ & \electromagnetic{} & \control{} & \yes & \no & \no & \yes & \yes \\
    Bayon et al. & 2016 & \cite{bayon_fault_2016} & TRNG & $\SI{0.003}{\watt}$ & $\SI{100}{\textcolor{black}{\textbf{\micro}}\meter}$ & \electromagnetic{} & \bias{} & \yes & \no & \no & \no & \no \\
    Park et al. & 2016 & \cite{park_pump_2016} & Medical & $\SI{0.030}{\watt}$ & $\SI{12.00}{\meter}$ & \optical{} & \bias{} & \no & \no & \no & \yes & \yes \\
    Roy et al. & 2017 & \cite{roy_backdoor_2017} & Microphone & $\SI{2.000}{\watt}$ & $\SI{1.50}{\meter}$ & \acoustic{} & \theoretical{} & \no & \yes & \no & \no & \no \\
    Song and Mittal & 2017 & \cite{song_inaudible_2017} & Microphone & $\SI{23.700}{\watt}$ & $\SI{3.54}{\meter}$ & \acoustic{} & \control{} & \no & \yes & \no & \no & \no \\
    Esteves and Kasmi & 2018 & \cite{esteves_remote_2018} & Microphone & $\SI{0.500}{\watt}$ & $\SI{10.00}{\meter}$ & \conducted{} & \control{} & \yes & \yes & \yes & \no & \yes \\
    Osuka et al. & 2018 & \cite{osuka_em_2019} & TRNG & $\SI{0.331}{\watt}$ & $\SI{0.40}{\meter}$ & \conducted{}\electromagnetic{} & \bias{} & \yes & \no & \no & \no & \no \\
    Roy et al. & 2018 & \cite{roy_inaudible_2018} & Microphone & $\SI{6.000}{\watt}$ & $\SI{7.62}{\meter}$ & \acoustic{} & \control{} & \no & \yes & \no & \no & \yes \\
    Selvaraj et al. & 2018 & \cite{selvaraj_induction_embedded_2018} & GPIO & $\SI{1.820}{\watt}$ & $\SI{1.00}{\meter}$ & \electromagnetic{} & \disruption{} & \yes & \yes & \yes & \yes & \no \\
    Giechaskiel et al. & 2019 & \cite{giechaskiel_framework_2019} & ADC & $\SI{0.010}{\watt}$ & $\SI{0.05}{\meter}$ & \conducted{}\electromagnetic{} & \theoretical{} & \yes & \yes & \yes & \yes & \yes \\
    Tu et al. & 2019 & \cite{tu_trick_2019} & Temp. \& Amp. & $\SI{3.162}{\watt}$ & $\SI{6.00}{\meter}$ & \conducted{}\electromagnetic{} & \control{} & \yes & \yes & \yes & \yes & \no \\
    Yan et al. & 2019 & \cite{yan_feasibility_2019} & Microphone & $\SI{1.500}{\watt}$ & $\SI{19.80}{\meter}$ & \acoustic{} & \control{} & \no & \yes & \no & \no & \yes \\
    \midrule
    Son et al. & 2015 & \cite{son_rocking_2015} & MEMS & $\SI{113}{\deci\bel}$ & $\SI{0.10}{\meter}$ & \acoustic{} & \disruption{} & \yes & \no & \no & \yes & \no \\
    Trippel et al. & 2017 & \cite{trippel_walnut_2017} & MEMS & $\SI{110}{\deci\bel}$ & $\SI{0.10}{\meter}$ & \acoustic{} & \control{} & \yes & \yes & \yes & \yes & \yes \\
    Zhang et al. & 2017 & \cite{zhang_dolphin_2017} & Microphone & $\SI{125}{\deci\bel}$ & $\SI{1.75}{\meter}$ & \acoustic{} & \control{} & \no & \yes & \no & \no & \yes \\
    Bolton et al. & 2018 & \cite{bolton_blue_2018} & HDD & $\SI{130}{\deci\bel}$ & $\SI{0.10}{\meter}$ &
    \acoustic{} & \disruption{} & \yes & \no & \no & \yes & \yes \\
    Shahrad et al. & 2018 & \cite{shahrad_acoustic_2018} & HDD & $\SI{103}{\deci\bel A}\hspace{-0.68em}$
    & $\SI{0.70}{\meter}$ & \acoustic{} & \disruption{} & \yes & \no & \no & \yes & \no \\
    Tu et al. & 2018 & \cite{tu_injected_actuation_2018} & MEMS & $\SI{135}{\deci\bel}$ & $\SI{7.80}{\meter}$ & \acoustic{} & \control{} & \yes & \no & \yes & \yes & \yes \\
    Khazaaleh et al. & 2019 & \cite{khazaaleh_vulnerability_2019} & MEMS & $\SI{94}{\deci\bel}$ & $\SI{0.11}{\meter}$ & \acoustic{} & \disruption{} & \yes & \no & \no & \yes & \yes \\
    \bottomrule
    \end{tabular}
    \label{table:attacks}
\end{table*}

Figure~\ref{fig:attacks} reveals that cross-influences are not limited
to attacks. Instead, there is a general trend of earlier research observing
the effects of non-adversarial interference, with later work actively
exploiting the same phenomenon for signal injection attacks. For example,
multiple works had identified the effects of electromagnetic interference
on medical devices~\cite{irnich_interference_1996, hayes_interference_1997,
lee_clinically_2009, seidman_vitro_2010} (with more papers
discussed in Section~\ref{sec:em}), but Foo Kune et al.~\cite{kune_ghost_2013}
were the first to recognize the effect as a security concern rather than
a safety and reliability one. Similarly, Dean et al.\ commented on the
effect of acoustic noise on gyroscopes~\cite{dean_degradation_2007,
dean_characterization_2011}, but Son et al.~\cite{son_rocking_2015} later
used the same effect to destabilize drones.

\begin{lesson}
    Out-of-band signal injection attacks identify the effect of noise
    on systems and find novel ways to amplify it through hardware imperfections.
\end{lesson}

The graph further shows that
out-of-band signal injection attack topics and methodology has both been
inspired by and has itself inspired research exploiting more
traditional avenues of attack. For example, investigations into acoustic signal injection
attacks~\cite{roy_backdoor_2017, zhang_dolphin_2017, roy_inaudible_2018}
have been influenced by research exploiting machine learning algorithms
which respond to commands which are audible but indecipherable by
humans~\cite{vaidya_cocaine_2015, carlini_hidden_2016}.
In addition, out-of-band acoustic attacks on
MEMS sensors~\cite{son_rocking_2015, trippel_walnut_2017}
have both inspired and been influenced by in-band optical attacks on
unmanned aerial vehicles (UAVs)~\cite{davidson_controlling_2016}.
Moreover, after the effects of acoustic noise on the
security of gyroscopes were first identified~\cite{son_rocking_2015},
subsequent work improved the level of control over the
output~\cite{tu_injected_actuation_2018}, and attacked additional types of
MEMS sensors~\cite{trippel_walnut_2017}, and
electro-mechanical devices such as HDDs~\cite{bolton_blue_2018}.

\begin{lesson}
    After an exploitable hardware imperfection has been identified, determining its
    root cause opens up new avenues of attacks across different domains, and
    with alternative methodologies.
\end{lesson}

Figure~\ref{fig:attacks} further highlights a few key works which in the opinion of the
authors have played a central role in the development of the field. The
first set of such papers~\cite{markettos_trng_injection_2009, bayon_trng_em_2012,
kune_ghost_2013, zhang_dolphin_2017, son_rocking_2015} was chosen due to the high number of citations
they have received overall ($\ge95$) and from other out-of-band attack
research ($\ge5$).\footnote{Citation counts are current as of 8 Nov. 2019
according to Google Scholar.}
Specifically, the works by Markettos and Moore~\cite{markettos_trng_injection_2009}
and by Bayon et al.~\cite{bayon_trng_em_2012} were chosen because they
successfully biased TRNGs in the conducted
and EM settings respectively, going beyond earlier theoretical work on
oscillator locking~\cite{adler_study_1946, mesgarzadeh_study_2005}. Their work
led to the development of a new branch of attacks, which has so far developed rather
independently of other out-of-band signal injections, as shown in Figure~\ref{fig:attacks}.

Foo Kune et al.'s work on adversarial electromagnetic interference~\cite{kune_ghost_2013}
also features prominently in Figure~\ref{fig:attacks}, having been cited
by almost all out-of-band signal injection attacks that were published after it
(with the exception of TRNG research). Foo Kune et al.\ were the first to successfully exploit
non-linearities and unintentional antennas in remote electromagnetic
injection attacks, which were non-adversarial in prior work
(e.g.,~\cite{irnich_interference_1996, hayes_interference_1997}) or only
mentioned in passing~\cite{rasmussen_proximity_based_2009}.

With over 160 citations since 2017, the research
by Zhang et al.~\cite{zhang_dolphin_2017} has been very influential
in the realm of out-of-band acoustic attacks against microphones. Unlike earlier work
on covert communication~\cite{roy_backdoor_2017}, and
indecipherable-yet-audible commands~\cite{vaidya_cocaine_2015, carlini_hidden_2016},
Zhang et al.'s ``DolphinAttack'' exploited microphone non-linearities
for inaudible injections.

Research on acoustic attacks is perhaps more mature against MEMS sensors,
in great part due to early work by Son et al., who first showed
how to disrupt gyroscopes~\cite{son_rocking_2015}. Moreover, Trippel et al.'s
research has also significantly furthered the state-of-the-art in acoustic
injections by controlling the output of accelerometers
for short periods of time~\cite{trippel_walnut_2017}. As a result,
the work of Tu et al., which showed
how to extend the duration of control~\cite{tu_injected_actuation_2018},
is in the second set of works highlighted
in Figure~\ref{fig:attacks}. This set contains recent studies whose novelty
and potential has not yet received mainstream attention.\footnote{Trippel et al's
2017 work~\cite{trippel_walnut_2017} lies between the two categories, already having over 75
citations, $12$ of which are from other attack papers.}
For example, the techniques proposed by Tu et al.\ to overcome
ADC sampling rate drifts should be applicable to other methods of injection, and
against different types of targets.
We place the work by Bolton et al.~\cite{bolton_blue_2018} in the same category,
as it managed to bridge research on HDD attacks (e.g.,~\cite{shahrad_acoustic_2018})
with attacks on MEMS sensors, and contained significant insights into why
resonance attacks against hard drives work.

The final work highlighted in Figure~\ref{fig:attacks} is the in-band attack
of Shoukry et al.\ against an Anti-lock Braking System (ABS)~\cite{shoukry_non_invasive_2013}.
It has been included not just for its high citation count ($>135$, of which
$10$ are out-of-band attacks), but because it is the first EM paper
to focus on the magnetic field rather than the electric field. As a result,
it serves as inspiration to recent out-of-band magnetic
attacks~\cite{selvaraj_induction_embedded_2018, nashimoto_sensor_2018},
which we hope will be explored more in the future (Section~\ref{sec:future}).

We also summarize the various out-of-band signal injection attacks
along with factors which contribute to them in Table~\ref{table:attacks}.
The table further
notes the maximum power used and distance achieved for an attack,
including the level of control over the resulting signal. Effects
range from theoretical attacks which are only partially realized to
practical attacks which disrupt, bias, or completely control the output.
As Table~\ref{table:attacks} indicates, information on the attack setup
was often hard to find, sometimes completely missing, and often had to
be identified by looking up the datasheets of the signal generators, antennas, and
amplifiers used. This lack of experimental details is further discussed
in the context of future research in Section~\ref{sec:future}.

We identify five key
aspects of vulnerability that attacks exploit: resonance; non-linearity;
improper filtering; poor shielding; and insecure algorithms. All attacks
which do not target microphones and optical sensors depend on
resonance of some sort: this can be acoustic resonance of mechanical structures,
electromagnetic resonant frequencies of unintentional antennas,
or the existence of locking frequencies for ring oscillators. Other attacks
depend on non-linearities of amplifiers, microphones, and speakers
to demodulate high-frequency signals. This is because resonant
frequencies are often much higher than those of desired injection waveforms.

In addition, many works identify improper filtering, particularly prior to
ADCs and amplifiers, as well as poor shielding as factors for
out-of-band attacks. Finally, in some cases, insecure
sampling and processing algorithms exacerbate the problem by making it easier
for an adversary to trick the system under attack into performing a dangerous
action. These sources of vulnerability form the basis for many
of the proposed countermeasures, which we discuss in detail
in Section~\ref{sec:defenses}.

\section{Analysis of Countermeasures}
\label{sec:defenses}

Although the literature on out-of-band attacks is quite broad,
research on defenses has been more sparse. Section~\ref{sec:oob_defenses} first summarizes
the state-of-the-art in countermeasures specific to out-of-band injections.
Section~\ref{sec:other_defenses} then expands our discussion by introducing general
protective and preventive approaches which remain applicable in this context.

\subsection{Out-of-Band Defense Mechanisms}
\label{sec:oob_defenses}

\begin{table*}[!t]
    \centering
    \caption{Summary of evaluated (\yes), proposed (\proposed), and absent (\no) countermeasures
             against \acoustic{}~acoustic, \conducted{}~conducted,
             \electromagnetic{}~electromagnetic, and \optical{}~optical out-of-band
             signal injection attacks.}
    \begin{tabular}{@{}rrrrccccccc@{}}
    \\\toprule
        \textbf{Authors} & \textbf{Year} & \textbf{Ref.} & \textbf{Target} & \rot{\textbf{Method}} &
        \rot{\textbf{Robust Hardware}}  & \rot{\textbf{Better Sampling}}    &
        \rot{\textbf{Sensor Fusion}}    & \rot{\textbf{Improved Filtering}} &
        \rot{\textbf{More Shielding}}   & \rot{\textbf{Anomaly Detection}}  \\
    \midrule
    Markettos and Moore & 2009 & \cite{markettos_trng_injection_2009} & TRNG & \conducted{}\electromagnetic{} & \proposed & \no & \no & \proposed & \proposed & \no \\
    Rasmussen et al. & 2009 & \cite{rasmussen_proximity_based_2009} & Medical & \electromagnetic{} & \no
    & \no & \no & \no & \proposed & \no \\
    Bayon et al. & 2012 & \cite{bayon_trng_em_2012} & TRNG & \electromagnetic{} & \no & \no & \no & \no & \no & \no \\
    Buchovecka et al. & 2013 & \cite{buchovecka_frequency_2013} & TRNG & \conducted{}\electromagnetic{} & \no & \no & \no & \no & \no & \no \\
    Kune et al. & 2013 & \cite{kune_ghost_2013} & Medical & \electromagnetic{} & \yes & \no & \no & \yes
    & \yes & \yes \\
    Kasmi and Esteves & 2015 & \cite{kasmi_iemi_2015} & Microphone & \electromagnetic{} & \proposed & \no & \no & \no & \proposed & \proposed \\
    Petit et al. & 2015 & \cite{petit_remote_2015} & Camera & \optical{} & \no & \no & \proposed & \proposed & \no & \no \\
    Shoukry et al. & 2015 & \cite{shoukry_pycra_2015} & Active Sensors & \electromagnetic{} & \no & \yes
    & \no & \no & \no & \no \\
    Son et al. & 2015 & \cite{son_rocking_2015} & MEMS & \acoustic{} & \proposed & \no & \no & \no & \yes & \no \\
    Bayon et al. & 2016 & \cite{bayon_fault_2016} & TRNG & \electromagnetic{} & \no & \no & \no & \no & \no & \no \\
    Cao et al. & 2016 & \cite{cao_exploring_2016} & TRNG & \conducted{} & \no & \no & \no & \no & \no & \no \\
    Park et al. & 2016 & \cite{park_pump_2016} & Medical & \optical{} & \no & \no & \no & \no & \proposed & \proposed \\
    Shin et al. & 2016 & \cite{shin_sampling_2016} & Medical & \optical{} & \no & \no & \proposed & \no & \proposed & \no \\
    Roy et al. & 2017 & \cite{roy_backdoor_2017} & Microphone & \acoustic{} & \no & \no & \no & \no & \no & \no \\
    Song and Mittal & 2017 & \cite{song_inaudible_2017} & Microphone & \acoustic{} & \no & \no & \no & \no & \no & \no \\
    Trippel et al. & 2017 & \cite{trippel_walnut_2017} & MEMS & \acoustic{} & \proposed & \yes & \no & \proposed & \proposed & \no \\
    Wang et al. & 2017 & \cite{wang_sonic_2017} & MEMS & \acoustic{} & \proposed & \no & \proposed & \no
    & \proposed & \proposed \\
    Zhang et al. & 2017 & \cite{zhang_dolphin_2017} & Microphone & \acoustic{} & \yes & \no & \no & \no & \no & \yes \\
    Bolton et al. & 2018 & \cite{bolton_blue_2018} & HDD & \acoustic{} & \proposed & \no & \proposed & \no & \yes & \yes \\
    Esteves and Kasmi & 2018 & \cite{esteves_remote_2018} & Microphone & \conducted{} & \no & \no & \no & \proposed & \no & \proposed \\
    Nashimoto et al. & 2018 & \cite{nashimoto_sensor_2018} & MEMS & \acoustic{}\electromagnetic{} & \no & \no & \yes & \no & \no & \no \\
    Osuka et al. & 2018 & \cite{osuka_em_2019} & TRNG & \conducted{}\electromagnetic{} & \no & \no & \no
    & \proposed & \proposed & \no \\
    Roy et al. & 2018 & \cite{roy_inaudible_2018} & Microphone & \acoustic{} & \no & \no & \no & \no & \no & \yes \\
    Selvaraj et al. & 2018 & \cite{selvaraj_induction_embedded_2018} & GPIO & \electromagnetic{} & \no &
    \no & \no & \proposed & \proposed & \no \\
    Shahrad et al. & 2018 & \cite{shahrad_acoustic_2018} & HDD & \acoustic{} & \proposed & \no & \no & \no & \proposed & \proposed \\
    Tu et al. & 2018 & \cite{tu_injected_actuation_2018} & MEMS & \acoustic{} & \no & \proposed & \proposed & \proposed & \proposed & \no \\
    Giechaskiel et al. & 2019 & \cite{giechaskiel_framework_2019} & ADC & \conducted{}\electromagnetic{}
    & \proposed & \proposed & \no & \proposed & \proposed & \proposed \\
    Khazaaleh et al. & 2019 & \cite{khazaaleh_vulnerability_2019} & MEMS & \acoustic{} & \proposed & \no
    & \no & \no & \proposed & \proposed \\
    Mahmoud and Stoilovi\'{c} & 2019 & \cite{mahmoud_timing_2019} & TRNG & \conducted{} & \no & \no & \no & \no & \no & \proposed \\
    Muniraj and Farhood & 2019 & \cite{muniraj_detection_2019} & Servo & \electromagnetic{} & \no & \yes
    & \no & \no & \no & \yes \\
    Tu et al. & 2019 & \cite{tu_trick_2019} & Temp. \& Amp. & \conducted{}\electromagnetic{} & \no & \no
    & \proposed & \proposed & \proposed & \yes \\
    Yan et al. & 2019 & \cite{yan_feasibility_2019} & Microphone & \acoustic{} & \yes & \no & \no & \no & \no & \yes \\
    Tharayil et al. & 2019 & \cite{tharayil_sensor_2019} & MEMS & \acoustic{}\electromagnetic{} & \no & \no & \yes & \no & \no & \yes \\
    Zhang and Rasmussen & 2020 & \cite{zhang_detection_2020} & Generic Sensors & \electromagnetic{} & \no & \yes & \no & \proposed & \proposed & \no \\
    \bottomrule
    \end{tabular}
    \label{table:defenses}
\end{table*}

The works that have investigated countermeasures against out-of-band
signal injection attacks have noted that a combination of prevention
and detection techniques both in software and in hardware are necessary
to improve security. We have divided the proposed defense mechanisms
into six categories: more resilient hardware; improved sampling algorithms;
sensor fusion and duplication; better filtering; additional shielding; and
anomaly detection of measurements and the environment. We discuss each
category in detail below, and summarize the various
proposals per paper in Table~\ref{table:defenses}. As the table indicates,
much of the discussion has been theoretical, with few works evaluating countermeasures
in practice.

\begin{lesson}
    The effectiveness of proposed countermeasures remains mostly theoretical, as
    practical implementations are often limited in scope, with superficial discussion
    of monetary and computational costs.
\end{lesson}

\secpar{Robust Hardware} In response to resonance and non-linearity
vulnerabilities, various works have
proposed preventive improvements in the hardware itself to make it more robust and
less susceptible to attacks. One of these improvements against electromagnetic
attacks reduces asymmetries in differential inputs to a system. By doing so,
attacker transmissions are received almost identically by the two
unintentional receiving antennas, and are severely attenuated.
For example, Markettos and Moore recommend reducing the
asymmetries in ring oscillators through ``carefully balanced transistors'',
or the use of differential ones, which are ``less affected by supply and
substrate noise''~\cite{markettos_trng_injection_2009}.
Similarly, Foo Kune et al.\ found that using differential rather than single-ended
comparators attenuated signals by up to $\SI{30}{\deci\bel}$~\cite{kune_ghost_2013}.
Although signals could still be injected, the power requirements to do so
increased significantly, thereby raising the bar for attackers.

Another approach is to change the sensors themselves, rather than attempt
to improve the physical layout of a circuit. For example, both
Shahrad et al.~\cite{shahrad_acoustic_2018} and Bolton et al.~\cite{bolton_blue_2018}
note that replacing Hard-Disk Drives (HDDs) with Solid-State Drives (SSDs)
thwarts acoustic resonance attacks due to a lack of moving parts. In a similar vein,
Zhang et al.\ noted that the iPhone 6 Plus
resisted their inaudible voice commands, since it is ``designed to suppress any
acoustic signals whose frequencies are in the ultrasound range''~\cite{zhang_dolphin_2017}.

Finally, better frontends with fewer non-linearities are less sensitive to
EMI noise~\cite{kasmi_iemi_2015, giechaskiel_framework_2019} and sonic
injections \cite{trippel_walnut_2017, wang_sonic_2017, zhang_dolphin_2017}.
They can therefore make it harder for adversaries to inject their desired
signals into the system. Such general designs are discussed in
greater detail in Section~\ref{sec:other_defenses}.

\secpar{Better Sampling} Many papers have proposed improvements in the sampling
technique to make it harder for an adversary to predict how a high-frequency
signal will be converted to a low-frequency one. In 2015, Shoukry et al.\
proposed an alternative method of sampling active sensors called ``PyCRA''
(for Physical Challenge-Response Authentication) to detect signal injection
attacks~\cite{shoukry_pycra_2015}. Active sensors ``perform some action to
evoke and measure a physical response from some measurable entity'', and include,
for instance, magnetic encoders measuring angular velocity. Shoukry et al.'s proposal
revolves around physical challenges to prove the absence of adversarial transmissions.
Specifically, when the actuator is off ({\em silenced}),
there should be no measured quantity unless an attack is taking place.
By only shutting down the actuator for a small period of time, PyCRA can detect attackers
without compromising the quality of sensor measurements and actuation results.
Adversaries cannot stop transmissions in time due to physical and computational
delay limits, allowing PyCRA to identify them~\cite{shoukry_pycra_2015}.
It should be noted that Shin et al.\
have suggested that PyCRA would require high computational overhead in practice,
both in terms of the minimum sampling rate needed to hit those physical limits,
and for the detectors themselves~\cite{shin_sampling_2016}. Moreover,
PyCRA requires active sensors, so it primarily protects against in-band attacks.

\begin{figure}[t!]
    \centering
    \includegraphics[width=\linewidth]{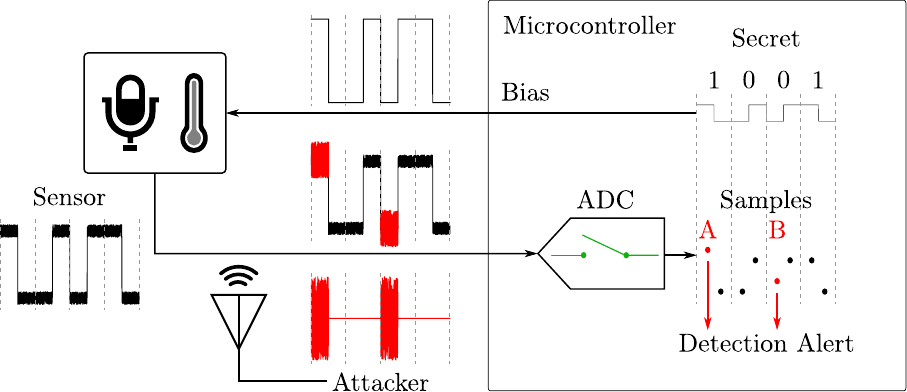}
    \caption{High-level overview of the defense mechanism by Zhang
             and Rasmussen~\cite{zhang_detection_2020}. Oversampling by a factor of
             $2n$ and selectively turning sensors on and off allows detection
             of out-of-band electromagnetic attacks: without knowing the secret
             sequence ($1001$ here for $n=4$), the adversary will cause inconsistent
             (A) or unexpected (B) non-zero samples.}
     \label{fig:defense}
\end{figure}

However, a similar proposal by Zhang and Rasmussen recently showed how to protect
both powered and non-powered passive sensors~\cite{zhang_detection_2020}. The key
idea is to use a secret bitstream to selectively turn off the sensor and observe
whether the measured signal has been altered by electromagnetic injections,
as shown in Figure~\ref{fig:defense}.
More concretely, for each sensor measurement, $2n$ ADC samples are taken, corresponding
to an $n$-bit secret sequence. Each secret bit is Manchester-encoded, so that
a 0-bit is represented as the pair $(0,1)$, corresponding to turning the sensors
off for one sample, and then turning them on (``biasing'') for another sample.
A 1-bit is similarly encoded as the pair $(1,0)$, first turning on the sensor,
and then turning it off during the second sample.

When the sensor is turned off, all samples should be close to zero, within some
noise- and device-dependent tolerance. Moreover, for fast-enough sampling
frequencies and slow-enough sensor signals, the $n$ samples when the sensor is
on should be close to each other. As a result, to inject a single measurement
successfully, an attacker needs to correctly predict the $n$ secret bits,
which only happens with probability $2^{-n}$ for a randomly chosen bit sequence.
By using a switch, non-powered passive sensors can also be adapted to use
this approach. Moreover, spikes in the frequency domain allow Zhang
and Rasmussen to detect attacker transmissions even for non-constant
sensor signals~\cite{zhang_detection_2020}. Overall, by oversampling
for each sensor measurement, noise can be distinguished from
adversarial signals with probabilistic guarantees.

An alternative approach to prevent attackers from injecting their desired
waveforms into a system is to add
randomness to the sampling process, especially for ADCs which
are only vulnerable for limited carrier frequencies~\cite{giechaskiel_framework_2019}.
The effect is essentially one of ``having an inaccurate ADC''~\cite{trippel_walnut_2017},
allowing a moving average to filter out injected periodic signals. This is similar
to sampling with a ``dynamic sample rate'', defeating the side-swing and
switching attacks of Tu et al.~\cite{tu_injected_actuation_2018}, which were explained in
Section~\ref{sec:acoustic}. Out-of-phase sampling has also been proposed
as a band-stop filter to reject frequencies near an accelerometer's resonant
frequency, thereby removing attacker-injected DC offsets~\cite{trippel_walnut_2017}.

It should be noted that protecting
against signal injection attacks into actuators has not been studied as extensively
in literature. However, Muniraj and Farhood recently proposed a
detection method based on a watermarking scheme that slightly alters the
actuation parameters~\cite{muniraj_detection_2019}, similar to
the proposal of Zhang and Rasmussen~\cite{zhang_detection_2020}.
Under attack, the measured response of the system does
not match the effect of the watermark, allowing detection. The same paper
also suggested pseudo-randomly changing the pulse frequency of the Pulse Width Modulation (PWM)
signals, making an actuator attack harder to accomplish. Overall, these methods
only alter the shape of the waveform that an adversary can inject rather than
the root cause of the vulnerability itself. They are therefore not
sufficient countermeasures by themselves.

\secpar{Sensor Fusion} A few works have suggested that using
sensors of different types ({\em fusion}) or multiple sensors of
the same type ({\em duplication}) with different vulnerable
frequency ranges will make injections harder~\cite{petit_remote_2015, shin_sampling_2016, wang_sonic_2017,
tu_injected_actuation_2018, bolton_blue_2018}. This is because an adversary
would need to mount multiple simultaneous attacks, which potentially
interfere destructively. However, in ``Sensor CON-Fusion'',
Nashimoto et al.~\cite{nashimoto_sensor_2018} showed that
a fusion algorithm based on Kalman filters could be circumvented (Section~\ref{sec:acoustic}).
As a result, better techniques are needed to protect against adversarial
injections, instead of simply faulty readings~\cite{park_sensor_2015, ivanov_attack_2016}.
Tharayil et al.\ proposed an improved such fusion algorithm, which takes into account
the mathematical relations between the underlying physical quantities. This allows them
to link measurements by a gyroscope and a magnetometer in a way which can
detect adversarial injections without any hardware modifications~\cite{tharayil_sensor_2019}.

Other researchers have proposed that additional sensors be used to measure and counteract
attacker signals. These additional components should not be identical to the
vulnerable sensors: as Khazaaleh et al.\ show, many MEMS gyroscopes
integrate a second, ``identical proof mass to perform differential measurements''
and ``eliminate unwanted vibrations''~\cite{khazaaleh_vulnerability_2019}.
However, they still remain vulnerable to ultrasonic attacks.

Instead, ``an additional gyroscope [...] that responds only to the resonant frequency''
may be able to remove the resonance effect from the main gyroscope~\cite{son_rocking_2015}.
Similarly, microphones might be able to detect (and potentially cancel) resonant frequencies
to protect MEMS gyroscopes~\cite{wang_sonic_2017} and HDDs~\cite{bolton_blue_2018}.
This approach would be hard in practice, at least for hard drives: the area to be
protected would need to cover ``the read/write head [completely] as it moves across the disk'',
and the sound wave to be generated would be potentially large,
raising many issues about its implementation~\cite{bolton_blue_2018}.

\secpar{Filtering} Most papers studied highlight the need for
better filters to reduce the vulnerable frequency range against
conducted~\cite{markettos_trng_injection_2009, esteves_remote_2018,
giechaskiel_framework_2019, tu_trick_2019},
electromagnetic~\cite{kune_ghost_2013, osuka_em_2019, selvaraj_induction_embedded_2018,
giechaskiel_framework_2019, tu_trick_2019, zhang_detection_2020},
acoustic~\cite{trippel_walnut_2017, tu_injected_actuation_2018},
and optical attacks~\cite{petit_remote_2015}. However, only Foo Kune et al.~\cite{kune_ghost_2013}
have performed systematic experiments studying the effect of filtering on the efficacy
of out-of-band signal injection attacks. To start with, Foo Kune et al.\ noted that adding
a low-pass filter in their experiments against bluetooth headsets allowed
audio signals to pass, but attenuated the injected electromagnetic signal
by $\SI{40}{\deci\bel}$. Moreover, they proposed an adaptive filtering
mechanism which uses the measured signal and the ambient EMI level to cancel
the attacker-injected waveform. Using a Finite Impulse Response (FIR) filter,
the algorithm estimates this waveform, and allows quick recovery of the
original signal, after an onset period at the beginning of the attack.
It should be noted, however, that filters might not be effective against
MEMS sensors: Khazaaleh et al.\ noted that ``false readings could not be
attenuated by adding a $\SI{10}{\hertz}$ low-pass filter'', despite
the resonant frequency being in the $\si{\kilo\hertz}$ range~\cite{khazaaleh_vulnerability_2019}.

\secpar{Shielding} Better separation from the environment also
improves protection against out-of-band signal injection attacks.
For this reason, it has been recommended by most authors investigating attacks
on sensors other than microphones. This shielding may come in the form of physical
isolation~\cite{son_rocking_2015, park_pump_2016, shahrad_acoustic_2018,
tu_injected_actuation_2018, khazaaleh_vulnerability_2019},
better acoustic dampening materials~\cite{trippel_walnut_2017, wang_sonic_2017,
tu_injected_actuation_2018, bolton_blue_2018}, or radio frequency
shielding~\cite{markettos_trng_injection_2009,
rasmussen_proximity_based_2009, kune_ghost_2013, kasmi_iemi_2015, shin_sampling_2016,
selvaraj_induction_embedded_2018, osuka_em_2019, giechaskiel_framework_2019,
tu_trick_2019, zhang_detection_2020}. For instance, Foo Kune et al.\
demonstrated a $\SI{40}{\deci\bel}$ attenuation of the injected signal,
even when the shielding had ``large imperfections''~\cite{kune_ghost_2013}.
These openings (e.g., for wires to pass through) result in ``major
degradations in the shielding''~\cite{tesche_emc_1996}. Indeed, Selvaraj et al.\ noted that
``while a light sensor can function in a mesh-based Faraday cage, magnetic shielding would
prevent light from reaching the sensor''~\cite{selvaraj_induction_embedded_2018}.
In addition, Bolton et al.\ showed that dampening foam ``significantly reduced
an HDD's susceptibility to write blocking'', but ``did not attenuate lower frequency
signals''~\cite{bolton_blue_2018}. Moreover, the foam led to an increased temperature
of $\SI{10}{\celsius}$, which can also result in disk failure.
As a result, ``it is often necessary to use a combination of shielding
and other protective measures''~\cite{tesche_emc_1996}.

\secpar{Anomaly Detection} Instead of trying to
prevent signal injection attacks, some works have proposed better software-level processing
of sensor signals, primarily for anomaly detection, with or without additional
hardware. One such approach is to estimate the ambient level of
electromagnetic~\cite{kune_ghost_2013, kasmi_iemi_2015, tu_trick_2019},
optical~\cite{park_pump_2016} and acoustic~\cite{wang_sonic_2017, zhang_dolphin_2017, blue_hello_2018,
bolton_blue_2018, shahrad_acoustic_2018, roy_inaudible_2018, yan_feasibility_2019}
emissions. For example, Park et al.\ noted that saturation attacks can be detected
simply ``by checking whether the light intensity exceeds the preset
maximum level''~\cite{park_pump_2016}. Foo Kune et al.~\cite{kune_ghost_2013} further investigated
the use of additional (intentional) antennas or reference conductors to measure
the levels of EMI radiation. This estimate can then be used by their adaptive
filtering algorithm~\cite{kune_ghost_2013}, which was discussed above.

In a similar vein, Tu et al.\ recommended the addition of a superheterodyne AM receiver to create
a tunable EM detector~\cite{tu_trick_2019}. This detector was shown to be useful
in estimating and compensating errors in the measurements. Other detection
mechanisms can operate with existing hardware: for example, Khazaaleh
et al.\ noted that ``sensing fingers'', which are already used to
measure displacement in the $y$-axis, can detect large displacements
caused by resonance~\cite{khazaaleh_vulnerability_2019}.

The question of how systems should behave once an attack has been detected has large
been side-stepped by many works. However, Bolton et al.\ introduced an algorithm
to augment the hard drive feedback controller and compensate for intentional
acoustic interference~\cite{bolton_blue_2018}. The addition of this attenuation
controller reduces the position errors of the read and write heads to
within the accepted tolerance levels, and allows the HDD to operate in the
presence of an attack.

Another way of detecting attacks is to use machine learning
classifiers~\cite{tharayil_sensor_2019}. However, such classifiers can be prone
to false positives, and will miss precise waveform injections. As a result,
it is often necessary to look for artifacts that would not
be present during the normal operation of a sensor, such as
harmonics and low or high frequency components~\cite{zhang_dolphin_2017,
blue_hello_2018, roy_inaudible_2018, giechaskiel_framework_2019,
yan_feasibility_2019}. This might not always be as straightforward as simply detecting
energy at low or high frequencies that are only present due to non-linearities:
for sophisticated attackers,
defense mechanisms need to exploit the properties of the legitimate signal
itself. For example, Roy et al.\ showed that ``voice signals
exhibit well-understood patterns of fundamental frequencies'', which are not
present in attacks and environmental noise. As a result, they can be used
to detect acoustic commands generated by ultrasound
signals~\cite{roy_inaudible_2018}. Similarly, the {\em absolute refractory period}
is hard for an attacker to spoof precisely via EM injections~\cite{kune_ghost_2013}.
This period represents the time span after a contraction during which the cardiac
tissue will not contract again. As shown by Foo Kune et al.~\cite{kune_ghost_2013},
it can be used to distinguish between a real and adversarial signals.

Finally, more restrictive processing of sensor data can also help mitigate
signal injection attacks. For example, safe defaults when the sensor output
is deemed as untrustworthy~\cite{kune_ghost_2013, park_pump_2016}
can reduce the effects of successful attacks on health- and safety-critical actions
taken by systems. Similarly, less permissive choices in the design
of voice interfaces can prevent non-targeted attacks from succeeding. As an example,
adding voice authentication and custom keywords can prevent command
injections into smartphones~\cite{kasmi_iemi_2015, esteves_remote_2018, giechaskiel_framework_2019}.

\begin{lesson}
    Until more resilient components replace vulnerable ones,
    defense-in-depth is necessary to
    protect against signal injection attacks. This can be accomplished
    through better filtering and shielding to prevent attacks, and through
    better sampling, fusion, and anomaly detection algorithms to identify them.
\end{lesson}

\subsection{Other Defensive Approaches}
\label{sec:other_defenses}

As out-of-band signal injection attacks are closely connected with
different areas of research (Section~\ref{sec:related}), there is extensive overlap
in the proposed countermeasures. For
example, before Foo Kune et al.~\cite{kune_ghost_2013} proposed an adversarial
EMI detector, Wan et al.~\cite{wan_increase_2010, wan_electromagnetic_2012}
introduced a similar design to ``increase the immunity of a microcontroller-based system in a
complex electromagnetic environment''. Moreover,
to protect against LiDAR attacks, Shin et al.\ proposed sensor fusion and redundancy,
fail-safe defaults, better shielding (by reducing receiving angle),
and randomized pinging directions and waveforms~\cite{shin_illusion_2017}.
Similarly, Davidson et al.\ proposed sensor fusion and an improved optical flow algorithm to
protect against optical in-band sensor spoofing~\cite{davidson_controlling_2016}.
Moreover, Blue et al.\ detected (audible) command injections by identifying
a frequency band which is produced by electronic speakers, but is
absent in human speech~\cite{blue_hello_2018}. In fact, detecting
unique features of the sensed property is a common defense mechanism for
general sensor manipulation attacks, such as those against Smart Grid power
plants~\cite{sridhar_model_based_2014, jiang_defense_2017}, or
unmanned aircraft systems~\cite{muniraj_detection_2019}. However,
such approaches require a theoretical system model, and assume an
adversary who cannot inject data obeying this model.

\begin{lesson}
    Defense mechanisms for in-band attacks, excessive environmental noise,
    and faulty sensors are often directly applicable to out-of-band signal
    injection attacks and vice versa.
\end{lesson}

In a different strand of research,
Redouté and Richelli have proposed some guidelines for improving immunity
against EM interference attacks~\cite{redoute_fundamental_2013, redoute_methodological_2015}.
These recommendations could be applied in the context of general out-of-band attacks:
\begin{enumerate}
\item {\em Filter induced signals before the non-linear device}.
      This suggestion is not limited to amplifiers,
      but can be used in other setups, including power transistors~\cite{calogero_new_2011}.
      It has been shown to result in
      an up to a $12.5\times$ reduction in EMI-induced offsets~\cite{walravens_efficient_2007,
      michel_differential_2011, redoute_fundamental_2013}, but may require
      bulky passive components, adding noise to the circuit.
\item {\em Linearize the stage generating the DC shift}, for example, by using
      amplifiers with a wider common mode input range, resulting
      in better linear behavior~\cite{sbaraini_emi_2010}.
\item {\em Prevent the accumulation of DC shift}, for instance by addressing
      the slew rate asymmetry and parasitic capacitances~\cite{richelli_robust_2004,
      fiori_design_2007, michel_comparison_2010}.
\item {\em Compensate and remove the induced offset}, for example,
      using cross-connected differential pairs~\cite{fiori_design_2007}.
\end{enumerate}
As discussing all possible EMI-resistant amplifier designs is out-of-scope,
the interested reader should refer to various comparative works~\cite{michel_comparison_2010, horst_emi_2014,
richelli_emi_2016} as a starting point. Similarly, one should refer
to advances in gyroscopic technologies~\cite{armenise_advances_2011, serrano_gyroscope_2016,
serrano_gyroscopes_2016} which do not use MEMS constructions, or reduce sensitivity
to random vibrations: as Khazaaleh et al.\ noted~\cite{khazaaleh_vulnerability_2019},
removing the ``misalignment between the sensing and driving axes'' will make
systems more secure against out-of-band acoustic attacks.

\begin{lesson}
    More accurate and sensitive hardware that is robust to environmental influences
    is a natural defense mechanism against out-of-band attacks.
\end{lesson}

\section{Additional Related Work}
\label{sec:related}

As this study contains the first survey of out-of-band signal injection attacks,
this section shows the close connections with side-channel
leakage and electromagnetic interference. For example, using
insights into the resonant frequencies of gyroscopes,
Farshteindiker et al.\ showed that unprivileged websites could act as
covert channel receivers, even at very low sampling frequencies of
$\SI{20}{\hertz}$~\cite{farshteindiker_phone_2016}. Block et al.\
improved the design by not requiring external equipment for the
attack, instead relying on the smartphone's speaker and
accelerometer~\cite{block_autonomic_2017}. Matyunin et al.\ then used the same
effect for cross-device tracking using ultrasonic transmissions at or
near the resonant frequencies of gyroscopes~\cite{matyunin_zero_2018}.
Moreover, Michalevsky et al.\ showed
that MEMS gyroscope measurements are sensitive to acoustic signals in their
vicinity~\cite{michalevsky_gyrophone_2014}. As a result, they
can be used to distinguish between different speakers, and, in part, the content
of the speech~\cite{michalevsky_gyrophone_2014} due to conducted vibrations of
the loudspeakers used~\cite{anand_speechless_2018, anand_spearphone_2019}.

In other words, the same source of vulnerability which can be used to
destabilize~\cite{son_rocking_2015} and control~\cite{trippel_walnut_2017, wang_sonic_2017,
tu_injected_actuation_2018} gyroscopes and accelerometers can be used for covert
channel communication~\cite{farshteindiker_phone_2016, block_autonomic_2017},
tracking~\cite{matyunin_zero_2018}, and speaker
identification~\cite{michalevsky_gyrophone_2014, anand_speechless_2018}.
Similarly, instead of using microphone non-linearities for command
injections~\cite{zhang_dolphin_2017, roy_inaudible_2018, yan_feasibility_2019},
Shen et al.~\cite{shen_jamsys_2019} and Chen et al.~\cite{chen_understanding_2019}
leveraged them to protect users' privacy by jamming nearby recording devices.

The countermeasures proposed in the works above mirror those of
Section~\ref{sec:oob_defenses}, and include anti-aliasing
filters, shielding, and sensor fusion.
Moreover, suggestions to increase noise in side- and regular-channel
emissions parallel out-of-band defense mechanisms based on reducing the
sampling accuracy. For example, decreasing ``the fidelity of the input
audio'' can prevent against inaudible voice injection
attacks~\cite{carlini_hidden_2016}. Similarly, fonts which
minimize emissions at high frequencies~\cite{kuhn_soft_1998,
tanaka_evaluation_2005, tanaka_information_2007} exploit the human eye
sensitivity to ``low spatial frequencies''~\cite{kuhn_soft_1998}.
As EM emanations of video display units (``TEMPEST'')~\cite{van_eck_electromagnetic_1985,
kuhn_electromagnetic_2004, kuhn_compromising_2013} mostly convey
``the high-frequency part of the video signal''~\cite{kuhn_soft_1998},
images are transformed in a way that is almost transparent to
human viewers, but prevents the reconstruction from side-channel listeners.
Researchers have likewise shown that adding certain patterns
to video frames~\cite{zhang_kaleido_2015} or the flashing of LED
lights~\cite{zhu_automating_2017} can reduce the fidelity of reconstructed images
from camera recordings, while not influencing regular viewers as much. Finally,
in some respects, anomaly detection resembles statistical and machine learning
approaches to detect covert channels, and can therefore draw inspiration
from seemingly unrelated disciplines, such as timing and storage
network covert channels~\cite{wendzel_pattern_2015}.

\begin{lesson}
    The sources of vulnerabilities for out-of-band signal injection attacks
    are often the same as those for hardware-based covert- and side-channel attacks.
    This allows the same techniques to be reused for attacks and defenses
    across disciplines.
\end{lesson}

Other research has indicated that devices which are typically used as actuators
can actually effectively function as sensors. LEDs can function as
photodiodes~\cite{loughry_oops_2019}, while speakers~\cite{guri_speakear_2017}
and HDDs~\cite{kwong_hard_2019} can both be converted into microphones.
Although all three attacks have so-far required the assistance of malware,
further research is required to identify the implications for out-of-band
signal injection attacks.

\begin{lesson}
    Reuse of off-the-shelf equipment in unconventional setups
    expands the surface for signal injection attacks
    exploiting hardware imperfections.
\end{lesson}

The lines between electromagnetic interference
and out-of-band signal injection attacks are also blurred. This is in part
because the self-classification of attacks depends
primarily on the research community with which an author is aligned,
rather than the end result of the injection. For example, the voice
injection command attacks of Kasmi and Lopes Esteves~\cite{kasmi_iemi_2015,
esteves_remote_2018} are categorized by their authors as Intentional
Electromagnetic Interference (IEMI) attacks, despite the relatively low power
used, and the lack of upsets or destruction of equipment. Similarly, Osuka
et al.\ considered their work to be in the IEMI realm~\cite{osuka_em_2019}, even though they biased
the randomness of a TRNG. This fact also partially explains why research
on out-of-band attacks against TRNGs has largely ignored attacks against other targets
and vice versa.

In general, this mismatch of expectations often results in
unexplored avenues of research, as can be seen, for example, in the IEMI attacks
on UAVs of Lopes Esteves et al.~\cite{esteves_unlocking_2018}: although
there is a strong inverse correlation between the battery temperature reading
and the strength of the electric field, the authors do not further investigate
how to precisely control the sensor output. Moreover, as was explained
in Section~\ref{sec:other_defenses}, research into electromagnetic interference
can provide insights into how to build more resilient hardware, even when
the hardware is only tested against
``unintentional parasitic signals and does not take a
malicious behavior of an attacker into account''~\cite{esteves_remote_2018}.

\begin{lesson}
    The proposed terminology based on the outcome rather than the method
    of injection can help systematize attack and defense approaches,
    and reveal previously unexplored connections.
\end{lesson}

\section{Future Directions}
\label{sec:future}

Despite the amount of research conducted on out-of-band signal injection
attacks, there is no common methodology to evaluate how susceptible systems are to
them. This is in contrast to related disciplines, such as
side-channel analysis~\cite{kocher_introduction_2011},
direct power injection and near-field scan immunity~\cite{boyer_modelling_2007},
fault injection attacks~\cite{yuce_fault_2018}, and IEMI
attacks~\cite{mansson_methodology_2009}.
Indeed, although many papers sweep through frequencies to
find the resonant ones~\cite{kune_ghost_2013, trippel_walnut_2017}, some do not
adopt this terminology~\cite{son_rocking_2015, bolton_blue_2018,
shahrad_acoustic_2018}, and do not specify how wide the frequency steps should be.
This proves to be particularly problematic, as some attack windows
``are as narrow as a few Hertz''~\cite{shahrad_acoustic_2018}.

Recently, Tu et al.~\cite{tu_injected_actuation_2018} provided a more detailed
methodology for acoustic injection attacks, which starts with a {\em profiling}
stage. During this phase, single-tone sounds are transmitted, and are swept at an interval of
$\SI{10}{\hertz}$. The devices targeted remain stationary during the profiling stage.
Further increments of
$\SI{1}{\hertz}$ or smaller can be used near the resonant frequencies to estimate
the sampling frequency of the ADC, and account for its drift. The next stage
involves {\em synchronizing} to a frequency which is close to a multiple of
the ADC sampling rate. This step is followed by {\em manipulating} the attack parameters,
and {\em adjusting} them in response to drifts. Although this approach
provides some common ground for evaluation, several
questions remain unanswered, especially when assessing countermeasures
to claim that a system is secure.
These questions include: what the frequency range itself should be; what the
step should be for wide ranges; what modulation method to use and with what
parameters; and whether there are other factors that
would need to be examined during experimentation. For instance, for electromagnetic
attacks, the incident angle of the EM field and the distance of attacks can have
a profound impact on their success, especially as they relate to
generalizing from the near- to the far-field.

\begin{lesson}
    A precise experimental procedure which specifies sweep, modulation, and
    other parameters is needed for out-of-band signal injection attacks.
\end{lesson}

The question of the maximum feasible attack distance
has mostly been of theoretical interest, with practical attacks often
limited to a few centimeters. Even though
EM attacks should in theory have a longer range than acoustic and optical attacks,
the converse appears to be true in the experiments conducted by the works
studied in this survey (Table~\ref{table:attacks}). There is also a worrying trend of assuming that more
power and more expensive equipment easily translates to a long-range attack.
For example, Tu et al.~\cite{tu_injected_actuation_2018} claim that with more speakers,
gyroscopes can be attacked from an $8\times$ longer distance, but
as Roy et al.~\cite{roy_inaudible_2018} showed, doing so is not a trivial engineering concern,
if the inaudibility of injections is to be maintained.

Similarly, although Foo Kune et al.~\cite{kune_ghost_2013}
claim that a $\SI{20}{\deci\bel}$ gain directional antenna
and a $\SI{1}{\watt}$ source can attack equipment at distances of up to $\SI{50}{\meter}$,
these estimates seem optimistic:
according to Lopes Esteves and Kasmi~\cite{esteves_remote_2018},
a $\SI{200}{\watt}$ source is required for a distance of $\SI{4}{\meter}$
for remote command injections~\cite{kasmi_iemi_2015}.
What is more, high-powered EM sources have the potential to cause faults in other
equipment and be harmful to human life. As a result, determining how to inject
precise signals from a distance is particularly challenging. These problems
become even more pronounced for magnetic attacks on actuators, which
have been limited so far.

\begin{lesson}
    Dedicated facilities and test equipment for
    long-range experimentation are needed.
\end{lesson}

Note that as many of the systems targeted are safety- and mission-critical,
we can expect that, in the future, some devices may be required to undergo
a certification process. Indeed, a CERT alert warning of
MEMS susceptibility to ultrasonic resonance~\cite{cisa_ics_2017} highlights that
out-of-band signal injection vulnerabilities are a concern for
governments and corporations alike. Regulations will thus pave the way for an expanding
industry around facilities and test equipment for EMC immunity against
adversarial injections.

As the currently published work often leaves experimental
details under-specified, reproducibility becomes a significant challenge:
for instance, as discussed in Section~\ref{sec:taxonomy}, details on the
power used were often not readily available, but required searching
through datasheets. Moreover, the duration of attacks was also often not specified.
Trippel et al.~\cite{trippel_walnut_2017} reported that some of their attacks
against accelerometers only work for a couple of seconds before the attack fails.
Sampling rate drifts thus necessitated manual tuning, or
more sophisticated attacking techniques, such as those proposed
by Tu et al.~\cite{tu_injected_actuation_2018}. However, without details
of the function generator specifications, it would be hard to know whether
some of the issues are caused by poor clock accuracy of the generator.
This problem is bound to become even more pronounced when using
Software-Defined Radio (SDR) and other low-end commodity hardware for
attack weaponization.

Minor variations in the construction of devices can also have significant effects
on the sensors' behavior, and will potentially impact the reproducibility of attacks.
For instance, Dey et al.~\cite{dey_accelprint_2014}
showed that otherwise identical accelerometers can be tracked
due to slightly different performance characteristics.
Giechaskiel et al.~\cite{giechaskiel_framework_2019} recently introduced security
definitions to address the lack of directly comparable metrics describing the
outcome of injection attacks. However, the overall
absence of experimental details, coupled with monetary
costs and legal requirements associated with using the electromagnetic spectrum,
make security research into out-of-band signal injection attacks a
challenging space for new researchers to enter.

\begin{lesson}
    Reproducibility through common metrics which allow for direct comparison
    of the effects of injection and standardized experimental
    setups are necessary to advance the state-of-the-art.
\end{lesson}

Besides the defense mechanisms of Section~\ref{sec:defenses},
to protect future devices for attack,
new security-sensitive products must take a fundamentally different approach
to trusting the outputs of sensors. In the words of Fu and Xu, there is a need
to ``shift from component-centric security to system-centric tolerance of
untrustworthy components'', perhaps taking note of advances in fault-tolerant
literature~\cite{fu_trusting_sensors_2018}. Fu and Xu also recommend that
sensor outputs be ``continuously checkable by software for adversarial influence'',
such as through internal debugging information that is hidden from accessible
APIs~\cite{fu_trusting_sensors_2018}. They further highlight the need for
interdisciplinary teams and education~\cite{fu_trusting_sensors_2018}. Indeed,
until new hardware is deployed, many cross-disciplinary solutions
will be required to prevent, detect, and mitigate attacks. As out-of-band
signal injection attacks become more powerful, collaboration will be necessary
to address the multifaceted research influences of the field.

\begin{lesson}
    Interdisciplinary research quantifying the effectiveness of
    countermeasures is needed to inform future hardware and
    software design choices.
\end{lesson}

\section{Conclusion}
\label{sec:conclusion}

Our ever-increasing reliance on sensors and actuators highlights the need for
a comprehensive look into electromagnetic, conducted, acoustic, and optical
out-of-band signal injection attacks. These attacks cause a mismatch between
a physical property being measured by a sensor or acted upon by an actuator
and its digitized version. Out-of-band signal injection attacks can
be used to control or disrupt drones, hard drives, and
medical devices, among others, with potentially
fatal consequences on human life. In light of the importance of
such attacks, this paper took the first step towards
unifying the diverse and expanding research through a taxonomy
of attacks, defenses, and terminology. Our work revealed
inter-disciplinary influences between seemingly disparate topics, and
also made several observations that can inform future research in the area.
Overall, better experimental and reporting procedures are necessary
for direct comparisons of the effects of attack and defense mechanisms.

\balance


\begin{IEEEbiography}[{\includegraphics[width=1in,height=1.25in,clip,keepaspectratio]{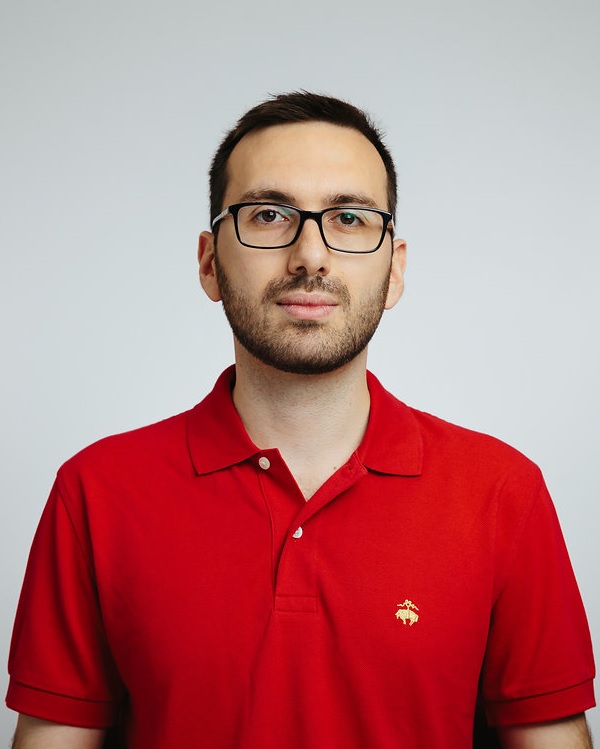}}]{Ilias Giechaskiel}
recently submitted his PhD thesis at the Department of Computer Science at the University
of Oxford. He was a Clarendon and Cyber Security Scholar at Kellogg College, and
was also funded by the EPSRC and the Oxford CDT in Cyber Security. His dissertation,
``Leaky Hardware: Modeling and Exploiting Imperfections in Embedded Devices''
was supervised by Prof. Kasper B. Rasmussen.
Ilias holds a Master's in Advanced Computer Science (with distinction) from the
University of Cambridge, and a Bachelor's in Mathematics (summa cum laude) from Princeton University.
During his PhD, Ilias was also a Visiting Assistant in Research at Yale University,
where he worked under Prof. Jakub Szefer on FPGA covert channels.
His interests in hardware security extend beyond academia: Ilias has participated
in numerous security Capture-the-Flag (CTF) competitions, and has interned at
Bloomberg, Microsoft, Dropbox, Microsoft Research, and Jump Trading.
\end{IEEEbiography}

\vspace*{-1.5em}
\begin{IEEEbiography}[{\includegraphics[width=1in,height=1.25in,clip,keepaspectratio]{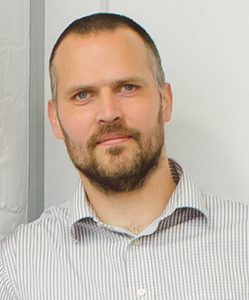}}]{Kasper B. Rasmussen}
is an Associate Professor in the Computer Science Department at the University of Oxford.
He joined the department in 2013 and was awarded a University Research
Fellowship from the Royal Society of London in 2015. Prior to being at Oxford, Kasper
spent two years as a post-doc at University of California, Irvine.
Kasper completed his Ph.D. with Prof. Srdjan Capkun in the Department of
Computer Science at ETH Zurich, where he worked on security issues
related to secure time synchronization and localization, with a particular
focus on distance bounding. His thesis won the ``ETH Medal'' for outstanding
dissertation from the Swiss Federal Institute of Technology, and he was additionally
awarded the Swiss National Science Foundation (SNSF) Fellowship for prospective researchers.
\end{IEEEbiography}

\end{document}